\documentstyle[12pt,epsf]{article}                                 
\input{epsf}

\begin{document}

\newcommand {\bea}{\begin{eqnarray}}
\newcommand {\eea}{\end{eqnarray}}
\newcommand {\be}{\begin{equation}}
\newcommand {\ee}{\end{equation}}

\newcommand{\pref}[1]{(\ref{#1})}

\def\IR{{\hbox{{\rm I}\kern-.2em\hbox{\rm R}}}}
\def\IH{{\hbox{{\rm I}\kern-.2em\hbox{\rm H}}}}
\def\IC{{\ \hbox{{\rm I}\kern-.6em\hbox{\bf C}}}}
\def\IZ{\relax\ifmmode\mathchoice
{\hbox{\sf Z\kern-.4em Z}}{\hbox{\sf Z\kern-.4em Z}}
{\lower.9pt\hbox{\sf Z\kern-.4em Z}}
{\lower1.2pt\hbox{\sf Z\kern-.4em Z}}\else{\sf Z\kern-.4em
}\fi}



%
%
\def\journal#1&#2(#3){\unskip, \sl #1\ \bf #2 \rm(19#3) }
\def\andjournal#1&#2(#3){\sl #1~\bf #2 \rm (19#3) }
\def\nextline{\hfil\break}

\def\ie{{\it i.e.}}
\def\eg{{\it e.g.}}
\def\cf{{\it c.f.}}
\def\etal{{\it et.al.}}
\def\etc{{\it etc.}}

\def\sst{\scriptscriptstyle}
\def\tst#1{{\textstyle #1}}
\def\frac#1#2{{#1\over#2}}
\def\coeff#1#2{{\textstyle{#1\over #2}}}
\def\half{\frac12}
\def\hf{{\textstyle\half}}
\def\d{\partial}

\def\inbar{\,\vrule height1.5ex width.4pt depth0pt}
\def\IR{\relax{\rm I\kern-.18em R}}
\def\IC{\relax\hbox{$\inbar\kern-.3em{\rm C}$}}
\def\IH{\relax{\rm I\kern-.18em H}}
\def\IP{\relax{\rm I\kern-.18em P}}
\def\Z{{\bf Z}}
\def\One{{1\hskip -3pt {\rm l}}}
\def\nth{$n^{\rm th}$}
%
%
\def\np#1#2#3{Nucl. Phys. {\bf B#1} (#2) #3}
\def\npb#1#2#3{Nucl. Phys. {\bf B#1} (#2) #3}
\def\pl#1#2#3{Phys. Lett. {\bf #1B} (#2) #3}
\def\plb#1#2#3{Phys. Lett. {\bf #1B} (#2) #3}
\def\prl#1#2#3{Phys. Rev. Lett. {\bf #1} (#2) #3}
\def\physrev#1#2#3{Phys. Rev. {\bf D#1} (#2) #3}
\def\prd#1#2#3{Phys. Rev. {\bf D#1} (#2) #3}
\def\annphys#1#2#3{Ann. Phys. {\bf #1} (#2) #3}
\def\prep#1#2#3{Phys. Rep. {\bf #1} (#2) #3}
\def\rmp#1#2#3{Rev. Mod. Phys. {\bf #1} (#2) #3}
\def\cmp#1#2#3{Comm. Math. Phys. {\bf #1} (#2) #3}
\def\cqg#1#2#3{Class. Quant. Grav. {\bf #1} (#2) #3}
\def\mpl#1#2#3{Mod. Phys. Lett. {\bf #1} (#2) #3}
\def\ijmp#1#2#3{Int. J. Mod. Phys. {\bf A#1} (#2) #3}
\def\jmp#1#2#3{J. Math. Phys. {\bf #1} (#2) #3}
\catcode`\@=11
\def\slash#1{\mathord{\mathpalette\c@ncel{#1}}}
\overfullrule=0pt
\def\AA{{\cal A}}
\def\BB{{\cal B}}
\def\CC{{\cal C}}
\def\DD{{\cal D}}
\def\EE{{\cal E}}
\def\FF{{\cal F}}
\def\GG{{\cal G}}
\def\HH{{\cal H}}
\def\II{{\cal I}}
\def\JJ{{\cal J}}
\def\KK{{\cal K}}
\def\LL{{\cal L}}
\def\MM{{\cal M}}
\def\NN{{\cal N}}
\def\OO{{\cal O}}
\def\PP{{\cal P}}
\def\QQ{{\cal Q}}
\def\RR{{\cal R}}
\def\SS{{\cal S}}
\def\TT{{\cal T}}
\def\UU{{\cal U}}
\def\VV{{\cal V}}
\def\WW{{\cal W}}
\def\XX{{\cal X}}
\def\YY{{\cal Y}}
\def\ZZ{{\cal Z}}
\def\lam{\lambda}
\def\eps{\epsilon}
\def\vareps{\varepsilon}
\def\underrel#1\over#2{\mathrel{\mathop{\kern\z@#1}\limits_{#2}}}
\def\lapprox{{\underrel{\scriptstyle<}\over\sim}}
\def\lessapprox{{\buildrel{<}\over{\scriptstyle\sim}}}
\catcode`\@=12

\def\sdtimes{\mathbin{\hbox{\hskip2pt\vrule height 5.5pt depth -.3pt 
width
.4pt \hskip-2pt$\times$}}}
\def\bra#1{\left\langle #1\right|}
\def\ket#1{\left| #1\right\rangle}
\def\vev#1{\left\langle #1 \right\rangle}
\def\det{{\rm det}}
\def\tr{{\rm tr}}
\def\mod{{\rm mod}}
\def\sinh{{\rm sinh}}
\def\cosh{{\rm cosh}}
\def\sgn{{\rm sgn}}
\def\det{{\rm det}}
\def\exp{{\rm exp}}
\def\sh{{\rm sh}}
\def\ch{{\rm ch}}

\def\lpl{\ell_{{\rm pl}}}
\def\lstr{l_s}
\def\str{{\sst\rm str}}
\def\bps{{\rm BPS}}
\def\ads{{\rm AdS}}
\def\inst{{\rm inst}}
\def\sltwoz{SL(2,\Z)}
\def\rhot{{\tilde\rho}}
\def\taut{{\tilde\tau}}
\def\Ptil{{\tilde P}}
\def\Ttil{{\widetilde \TT}}
\def\Jtil{{\widetilde \JJ}}
\def\Jtilbar{{\overline{\widetilde \JJ}}}
\def\extrat{{\tilde T}}
\def\ellt{{\tilde\ell}}
\def\gs{g_{\rm s}}
\def\gsix{g_6}
\def\nufour{\nu_4}
\def\adot{\dot a}
\def\Ytil{{X}}
\def\lamt{{\psi}}
\def\kth{$k^{\rm th}$}
\def\symn{{\sl Sym}^N}
\def\ghat{{\hat G}}
\def\hq{{\HH_{\vec q}}}
\def\eee{\varepsilon}

\def\bh{{\sst BH}}
\def\lpl{\ell_{{\rm pl}}}
\def\gym{g_{\sst\rm YM}}
\def\str{{\sst\rm str}}
\def\ttil{\tilde T}
\def\that{\hat T}
\def\symn{{\sl Sym}^N}
\def\adot{{\dot a}}
\def\bdot{{\dot b}}
\def\psibar{\bar\psi}

\begin{titlepage}
\rightline{EFI-99-39}

\rightline{hep-th/9909088}

\vskip 3cm
\centerline{\Large{\bf Currents and Moduli in the $(4,0)$ theory}}
\vskip 2cm
\centerline{
Finn Larsen\footnote{\texttt{flarsen@theory.uchicago.edu}}~~ {\it 
and} ~~
Emil Martinec\footnote{\texttt{ejm@theory.uchicago.edu}}}
\vskip 12pt
\centerline{\sl Enrico Fermi Inst. and Dept. of Physics}
\centerline{\sl University of Chicago}
\centerline{\sl 5640 S. Ellis Ave., Chicago, IL 60637, USA}
\vskip 2cm

\begin{abstract}
We consider black strings in five dimensions and their description 
as a $(4,0)$ CFT. The CFT moduli space is described 
explicitly, including its subtle global structure. 
BPS conditions and global symmetries determine the spectrum of 
charged excitations, leading to
an entropy formula for near-extreme black holes in four 
dimensions with arbitrary charge vector. In the BPS limit, this 
formula
reduces to the quartic $E_{7(7)}$ invariant. The prospects for a 
description of the $(4,0)$ theory as a solvable CFT are explored.
\end{abstract}

\end{titlepage}

\newpage


\section{Introduction and Summary}
The description of black holes as states in certain unitary conformal
field theories remains one of the highlights of recent progress in 
string 
theory (for reviews see 
{\it e.g.}~\cite{Maldacena:1997tm,Peet:1997es,Youm:1997hw}). 
The canonical setting for the discussion is the 1+1 dimensional
theory obtained from bound states of
D1-branes and D5-branes, with the D5 dimensions 
transverse to the D1 wrapped on a small four-manifold $T^{4}$ 
or $K3$~\cite{Strominger:1996sh}. This theory has $\NN=(4,4)$ 
supersymmetry 
and describes black strings in six dimensions; which compactify to 
black 
holes in five dimensions. It is clearly of interest to understand a 
broader 
class of theories which are similarly relevant for black holes. The 
next 
simplest example is a conformal field
theory with $\NN=(4,0)$ supersymmetry that describes black strings in 
five 
dimensions, or black holes in four 
dimensions. This theory is the subject of the present investigation, 
which may be considered a sequel to our previous work
on the D1-D5 system~\cite{lm99}; 
however, we will see that the
$(4,0)$ theory involves several interesting complications not present 
in the $(4,4)$ case.

In the AdS/CFT correspondence~\cite{adsrev} the $(4,4)$ theory is 
interpreted as string theory on 
$AdS_{3}\times S^{3}\times M$
~\cite{Hyun:1997jv,Boonstra:1997dy,Strominger:1998eq}, 
where $M=T^{4}$ or $M=K3$. The corresponding interpretation of the 
$(4,0)$ theory involves string theory on the orbifold 
$AdS_{3}\times S^{3}/\IZ_{N}\times M$~\cite{Kutasov:1998zh}; 
or M-theory on $AdS_{3}\times S^{2}\times 
X$~\cite{Balasubramanian:1998ee}, 
where $X$ is some Calabi-Yau three-fold, 
for which we exclusively take $X=T^{6}$.
The geometries underlying the $(4,4)$ and $(4,0)$
theories are therefore quite similar;
however, the $(4,0)$ theory is more challenging. First of all, the 
supersymmetry is reduced; so many properties of the theory are no 
longer constrained by general principles. More profoundly, the 
electric/magnetic duality, special to four dimensions, gives rise
to new structures that require special considerations. These 
interesting structures represent novel elements that have not 
previously 
been analyzed in detail.

There are several specific brane configurations that serve as explicit
examples of backgrounds, analogous to D1/D5 for the $(4,4)$ theory. 
One has three M5-branes intersecting over a common 
line~\cite{Klebanov:1996mh,Balasubramanian:1996rx}; another is a type 
IIB configuration with a fundamental string, a NS5-brane, and a 
KK-monopole~\cite{Cvetic:1996uj,Cvetic:1996yq}. 
These backgrounds are introduced in the beginning of 
section~\ref{back} 
along with more general families, depending on up to $9$ background 
charges. 
We use all of these examples repeatedly throughout the paper.

The maximally supersymmetric vacuum in five dimensions depends on $42$
moduli, parametrizing the coset 
$E_{6(6)}/USp(8)$~\cite{Cremmer:1979up}. 
The backgrounds we consider break some of the supersymmetry, as 
discussed 
above, and also restrict the moduli space by fixing the values of 
some of 
the moduli in 
terms of the background charge vector. 
In section~\ref{fixed} we discuss 
this mechanism in detail, giving the fixed values of the moduli 
explicitly, and enumerating the moduli that remain undetermined. 
There are $28$ such ``free'' moduli, parametrizing the coset
$F_{4(4)}/SU(2)\times USp(6)$
~\cite{Andrianopoli:1997hb,Lu:1997bg,Kutasov:1998zh}.

This discussion determines the {\it local} geometry of the moduli 
space of the CFTs, but the {\it global} structure requires
more care~\cite{Seiberg:1999xz,lm99}. In the $(4,4)$ theory the 
essential subtlety is captured by a subspace of moduli space which is 
locally isomorphic to the coset $SL(2,\IR)/U(1)$, 
which can be parametrized by the complex IIB string coupling.
In this subspace the global identifications are described by the 
group 
$\Gamma_{0}(N=q_{1} q_{5})$, acting on the moduli space from 
the left~\cite{lm99}. These identifications form a 
genuine subgroup of the $SL(2,\IZ)$ identifications
under IIB S-duality.  In section~\ref{glob}, we develop the 
corresponding
description for the $(4,0)$ theory. We find that the essential 
structure of 
moduli space is captured by the five dimensional coset 
$SL(3,\IR)/SO(3)$. 
Again, the global identifications are a genuine subgroup of the 
na\"{\i}ve 
expectation $SL(3,\IZ)$. Other features of the moduli space are also 
described explicitly in terms of the $SL(3,\IR)/SO(3)$ subspace of 
moduli
space.

A key feature of our discussion is the scaling limit introduced in 
section~\ref{scaling}. In this limit the $(4,0)$ theory decouples 
from 
the ambient environment~\cite{Maldacena:1997re}. 
An important distinction arises 
between the branes that are interpreted as sources of the background, 
and those that are charged excitations. The spectrum of the charged 
excitations is one of our main interests; it is the topic of 
section~\ref{exc}. We find a total of $27$ charges. A part of their 
spectrum is similar to that of perturbative winding/momentum charges 
on three independent four-tori; altogether this sector describes $24$ 
charges. 
The remaining $3$ charges are the ``electric'' duals of 
the three ``magnetically charged''
branes that are the sources of the background. These $3$ 
charges are 
special to the $(4,0)$ theory, and their spectrum is more 
complicated. 
We find their conformal weights using 
a combination of BPS algebra and global symmetries. The
results are then used in section~\ref{bhcft} to compute the entropy
of a large class of near-extremal black holes, depending on an 
arbitrary
charge vector in addition to independent mass and angular momentum 
parameters. The corresponding classical black holes have not yet been 
constructed in general, and their area is not known. However, our 
formula 
for the entropy reduces in the BPS-limit to the quartic invariant 
of $E_{7(7)}$, as it should. This result provides a fairly intricate 
test 
of our microscopic description.

One of the goals of the present investigation is to learn more about
the full CFT underlying the system. It is known that the $(4,4)$ 
theory 
is described in a region of moduli space by a solvable CFT, namely a 
sigma-model with a symmetric product orbifold as target 
space~\cite{Strominger:1996sh}. The $(4,0)$ theory is understood
in significantly less detail. For example, it is not known whether 
there
is an underlying solvable CFT and, if so, in what region of moduli
space it would be applicable. In the BPS limit many properties are
constrained from general principles and a formula has been proposed
for the spectrum~\cite{Dijkgraaf:1997it}; however, many features 
remain mysterious. We find several new facts about the theory which
may be helpful in identifying a solvable limit. In section~\ref{cft}
we explore the possibility of a description in terms of a relatively
modest variation on the symmetric product idea. Our results are
consistent with this working assumption; however, we have not 
succeeded
in determining the precise theory.

The paper is organized as follows. In section~\ref{scaling} we 
describe the scaling limit defining a theory decoupled from
gravity. This leads to the fundamental distinction between branes that
are considered part of the background, and those that are excitations.
The following sections~\ref{back} and \ref{exc} describe properties
of the background and the excitations, respectively. Each of these
sections has a fairly
large number of subsections, focussing on specific features. In
section~\ref{cft}, we discuss our attempts towards a description of
the $(4,0)$ theory as a solvable CFT. Finally, Appendix~\ref{app:BPS}
contains the computation of the basic spectrum of excitations.


\section{The Scaling Limit}
\label{scaling}
We consider four dimensional vacua 
with maximal supersymmetry; in other words M-theory on $T^{7}$.
This theory has $56$ $U(1)$ charges: $21$ from wrapped M5-branes, 
$21$ 
from wrapped M2-branes, $7$ KK-momenta, and $7$ KK-monopoles. The 
$M2/M5$ branes and the KK-momenta/KK-monopoles form dual pairs under
electric/magnetic duality in four dimensions.
 
We are interested in a scaling limit where a theory without gravity 
decouples form the bulk. The limit is most conveniently introduced
on a rectangular torus where it can be defined as~\cite{juanads}:
\be
l_p\to 0~~~{\rm with}~~R_{1},\cdots,R_{6}\sim l_p~~~~R_7\sim 1~.
\ee
The undemocratic treatment of the toroidal radii introduces a 
hierarchy among the $U(1)$ charges which is of central importance for 
our 
considerations. This is seen by inspecting the masses of isolated 
objects 
that carry each of the $56$ $U(1)$  charges\footnote{The precise 
definitions of the units are $l_p = g^{1/3}l_s$ where 
$l_s=\sqrt{\alpha^\prime}$.}:
\bea
M_{M5} &=& Q_{ij} ={1\over 5!}
\epsilon_{ijklmno}{R^{k} R^{l} R^{m} R^{n} R^{o}\over l^6_p}~, 
\label{m5mass}\\
M_{M2} &=& Z^{ij}={R^{i}R^{j}\over l^3_p}~, 
\label{m2mass}\\
M_{KK} &=& P^{i}={1\over R_i}~,
\label{kkmass}\\
M_{KKM} &=& {\tilde P}_{i}=
{R_i R_1 R_2 \cdots R_7\over l^9_p}~.
\label{kkmmass}
\eea
In the scaling limit the most massive excitations have mass
of order $l_p^{-3}$; they are the KK monopoles with KK direction 
along 
the large dimension $R_7$. The $M2$- and $M5$-branes wrapping $R_7$, 
and 
the KK-monopole not wrapping $R_7$ have masses of order $l_p^{-2}$; 
there are $6+15+6=27$ such excitations. The $M2$-branes and 
$M5$-branes not 
wrapping $R_7$ and the KK momentum along any direction except $R_7$ 
have 
masses of order $l_p^{-1}$; there are also $6+15+6=27$ of these
--- as there must be, by electric/magnetic duality. Finally, the KK 
momentum 
along $R_7$ has finite mass.

It is sometimes useful to consider the effective five-dimensional 
theory which corresponds to avoiding compactification along the 
``large''
dimension $R_7$ altogether. In this theory the superheavy KK 
monopoles 
(with $M\sim l^{-3}_p$) do not exist; this is one of the reasons that 
we do 
not consider these objects in this work. The excitations with mass of 
order 
$l^{-2}_p$, $l^{-1}_p$, and $l^0_p$ correspond to strings, particles, 
and 
waves in the effective five-dimensional theory. They transform under 
the 
five-dimensional duality group $E_{6(6)}$ as ${\bf 27}$, ${\bf 27}$, 
and 
${\bf 1}$. The reason that the hierarchy is fundamental for our 
considerations is that the objects enter very differently in the 
microscopic description: the excitations with mass of order 
$l^{-2}_p$, $l^{-1}_p$, and $l^0_p$ are interpreted as background 
fields, charged excitations in the background theory, and neutral
chiral excitations. Section~\ref{back} is devoted to a 
discussion of the background, and section~\ref{exc} considers the 
remaining
$U(1)$ charges, as excitations of the theory governing the background.

The discussion above was for rectangular tori. A more general set of 
moduli can be restored without further complications, as long as the
effective five-dimensional nature of the configuration is respected.
This allows for $21$ components of a general metric $G_{ij}$ 
($i,j,\cdots =1,\cdots 6$), $20$ components of the three-form field 
$C_{ijk}$ ($i,j,\cdots =1,\cdots 6$), and the pseudo-scalar 
${\cal E}_{123456}$.
(${\cal E}_{IJKLMN}$ is obtained as the potential 
dual of the 3-form field $C_{JKL}$.) 
These $42$ scalars parametrize the coset
space $E_{6(6)}/USp(8)$, as expected in a maximally supersymmetric
vacuum in five dimensions. 

\section{The Background}
\label{back}
The background is chosen such that it preserves precisely 1/8 of SUSY.
This condition ensures that the effective theory after decoupling is 
a 
conformal field theory (CFT) in $1+1$ dimensions, and also that 
excitations 
of the background correspond at strong coupling to regular black 
holes 
in $3+1$ dimensions. Duality of supergravity ensures that all 
backgrounds 
preserving 1/8 SUSY are classically equivalent; in fact, they all 
correspond 
to the near-horizon geometry $AdS_3\times S^2\times T^{6}$. That 
this classical symmetry generalizes to the full quantum description 
is part of the conjecture embodied in M-theory.

\subsection{The Canonical Backgrounds}
\label{can}
In explicit computations it is useful to choose a specific background.
One example is three $M5$-branes that intersect over a string, with 
the 
string aligned along the ``large'' dimension 
(of length $R_7$)~\cite{Klebanov:1996mh,Balasubramanian:1996rx}:
\newcommand{\bin}{$\bullet$}
\newcommand{\bout}{--}
\begin{center}
\begin{tabular}{c|ccccccc} 
        Brane  &  1   & 2    &3     &4     &5     &6      & 7    \\ 
\hline
        $M5$ & \bin & \bin & \bin & \bin & \bout & \bout & \bin   \\ 
        $M5$ & \bout & \bout & \bin & \bin & \bin & \bin & \bin   \\ 
        $M5$ & \bin & \bin & \bout & \bout & \bin & \bin & \bin  
\end{tabular}
\end{center}
An advantage of this representation is that it exhibits many 
symmetries
of the situation explicitly. For example, a triality of the three 
background objects is immediately apparent. This background will be 
our 
main example.

Another example is a type IIB string configuration with a
KK-monopole, an $NS5$ brane, and a perturbative string $F1$. 
It is~\cite{Cvetic:1996uj,Cvetic:1996yq}:
\begin{center}
\begin{tabular}{c|cccccc} 
        Brane &  1   & 2    &3     &4     & 5   &6        \\ \hline
        $KKM$ & \bin & \bin & \bin & \bin & $\times$ & \bin  \\ 
        $NS5$ & \bin & \bin & \bin & \bin & \bout & \bin     \\ 
        $F1$  & \bout & \bout & \bout & \bout & \bout & \bin     
\end{tabular}
\end{center}
where the $\times$ denotes the Taub-NUT direction of the KK-monopole. 
An  advantage of this representation is its close relation to the 
$D1/D5$ system. (The $F1/NS5$ in the background becomes
$D1/D5$ after type IIB S-duality.) 
Moreover, it is purely NS; so one may consider worldsheet 
string theory in this background~\cite{Kutasov:1998zh}. 
Note that in the type IIB representation it is $R_{6}$ which is 
``large''.

The two examples given above are related by the duality chain:
\be
M \stackrel{6-7~{\rm flip}}{\longrightarrow} M
\stackrel{M_{\rm red}~{\rm on}~7}{\longrightarrow} 
	IIA \stackrel{T_{125}}{\longrightarrow}IIB 
\stackrel{S}{\longrightarrow}IIB~.
\label{eq:duality}
\ee
Results obtained in one representation can be translated to the other
using this sequence of transformations.

\subsection{More General Backgrounds}
\label{sec:moreg}
It is often important to consider more general backgrounds. 
The most general case with all 27 background branes turned on is 
unfortunately quite complicated. In this section we introduce
an intermediate situation where one can turn on 9 different 
background branes, while still keeping things explicit. The
construction will be exploited repeatedly in the sequel.

The idea is that $E_{6(6)}$ has a maximal subgroup 
$SL(3)\times SL(3)\times SL(3)$, with the $\bf 27$ decomposing as 
$(3,3,1)+(1,3,3)+(3,1,3)$.  These three sets of background
charges can be thought of as matrices:
\be\label{oddcharge}
 {\bf Q_o}=\pmatrix{
 z_{17} & z_{37} & z_{57} \cr
 q_{35} & q_{15} & q_{13} \cr
 {\tilde p}_1 & {\tilde p}_3 & {\tilde p}_5} \ ,
\label{eq:char1} 
\ee
\be\label{evencharge}
 {\bf Q_e}=\pmatrix{
 z_{27} & z_{47} & z_{67} \cr
 q_{46} & q_{26} & q_{24} \cr
 {\tilde p}_2 & {\tilde p}_4 & {\tilde p}_6 }\ ,
\label{eq:char2} 
\ee
as well as:
\be\label{fivecharge}
{\bf Q_5}=\pmatrix{q_{12}& q_{32}&q_{52}\cr
q_{14}& q_{34}&q_{54}\cr
q_{16}& q_{36}&q_{56}}~.
\label{eq:char3} 
\ee
In these formulae the $z_{ij}$ are the integer number
of M2-branes wrapping the corresponding cycles; the ${\tilde p}_i$
are KK-monopoles wrapping the compact space,
with $i$ the monopole circle; and the $q_{ij}$ are M5-branes
with the indices reffering to the dual cycle on the $T^{7}$.
The $E_6$ cubic invariant specializes to the sum of the
determinants of these three matrices.

The charge matrices (\ref{eq:char1}-\ref{eq:char3}) 
are related by discrete symmetries.
First, the exchange of even and odd subspaces
interchanges ${\bf Q_o}$ and ${\bf Q_e}$; and takes 
${\bf Q_{5}}\longrightarrow {\bf Q_{5}^{t}}$.
Next, the sequence of dualities:
\be\label{5oflip}
M \stackrel{M_{\rm red}~{\rm on}~6}{\longrightarrow} 
	IIA \stackrel{T_{1345}}{\longrightarrow}
IIA \stackrel{M_{\rm lift}~{\rm on}~6}{\longrightarrow} M~,
\label{eq:triality}
\ee
interchanges ${\bf Q_o}$ and ${\bf Q_5}$; and takes
${\bf Q_{e}}\to \Gamma {\bf Q_{e}^{t}}\Gamma$, where:
\be
\Gamma=\pmatrix{& & 1\cr &1& \cr 1& & }~.
\ee
Taken together, these tranformations generate a triality map between 
any two sets of nine charges above.

Under the decomposition $E_{6(6)}\rightarrow SL(3)^3$, the maximal 
compact subgroup decomposes as $USp(8)\rightarrow SO(3)^3$. A 
conveniently 
described subspace of moduli space is therefore $[SL(3)/SO(3)]^3$.  
In the present construction we focus on these $15$ moduli, and turn
the remaining ones off. This restriction is such that explicit 
computations
remain possible, with results that are representative of the general 
case with $42$ moduli.

Two of the three $SL(3)/SO(3)$ cosets are the unit volume metrics 
${\hat G}_{2i,2j}$ and ${\hat G}_{2i-1,2j-1}$ on the three-subtori 
of even and odd cycles. 
These metrics are conveniently written using vielbeins
as ${\hat G}=e^t\cdot e$, for which we choose an $SL(3)/SO(3)$ coset 
representative of the form $e=A\cdot N$, where $A$ is diagonal with 
unit
determinant and $N$ is upper triangular with unit diagonal.
Henceforth we will drop the hats, remembering that the metrics
are of unit normalization. The last $SL(3)/SO(3)$ subspace is 
parametrized by the volumes $V_{135}$, $V_{246}$ 
and constant three-form fields $C_{135}$, $C_{246}$ on the even
and odd subtori, as well as the six-form modulus $\EE_{123456}$.
The vielbein parametrization is also convenient for describing this 
part of the moduli space, for which we may write:
\be\label{fivebein}
  e_5=(V_{246}/V_{135})^{1/6}\times
	\pmatrix{\sqrt{\coeff{1}{V_{135}V_{246}}}& & \cr
	 & \sqrt{\coeff{V_{135}}{V_{246}}}& \cr
	 & &\sqrt{{\scriptstyle V_{135}V_{246}}} }
    \pmatrix{1 & C_{135} & \EE_{123456}^{\prime} \cr
	 & 1 &  C_{246} \cr
	 & & 1 }\ ,
\ee
with the `metric' $G_5=e_5^t\cdot e_5$. We introduced
$\EE_{123456}^{\prime}=\EE_{123456}-{1\over 2}C_{135}C_{246}$.

The discrete symmetries also act simply on the moduli.
The exchange of odd and even cycles interchanges $G_{o}$ and $G_{e}$; 
and transforms $G_{5}$ as $G_{5}\rightarrow \Gamma G_{5}^{-1}\Gamma$. 
The sequence of dualities (\ref{eq:triality}) inverts the odd metric
$G_{o}\rightarrow G_{o}^{-1}$ and interchanges $G_{e}$ and $G_{5}$
according to $G_{e,5}\rightarrow 
\Gamma G_{5,e}\Gamma$.

\subsection{The Mass of the Background}
\label{sec:masssec}
The mass formula for a general background is some $E_{6(6)}$ invariant
combination of the charges which is necessarily quite complicated. 
However, 
in the various special cases considered in sections 
(\ref{can}-\ref{sec:moreg})
the details can be 
carried 
out explicitly. The most general mass formula follows in 
principle, by acting on the charges and the moduli with dualities.

We first consider the canonical $M5$-brane background. On a 
rectangular torus 
the 
mass of is simply $M = Q_{12}+Q_{34}+Q_{56}$, the sum of the 
constituent 
masses; but in general the mass depends nontrivially on the moduli. 
In fact, it is a general phenomenon that, in the presence of any
configuration of branes, the parity-odd moduli induce additional 
charges 
that are not na\"{\i}vely present. For a general four-dimensional 
configuration 
this effect can be taken into account by the 
shifts~\cite{Obers:1998fb}:
\bea
{\hat P}_I &=& P_I + {1\over 2}C_{IJK}Z^{JK}
+ ({1\over 4!}C_{JKL}C_{MNI}+{1\over 5!}{\cal E}_{JKLMNI})
Q^{JKLMN}~,
\label{eq:mshift1}\\
{\hat Z}^{IJ} &=& Z^{IJ} + {1\over 3!}
C_{KLM}Q^{IJKLM} + \nonumber \\
&~&\qquad + ({1\over 4!}C_{KLM}C_{NPQ}+ {1\over 5!}{\cal E}_{KLMNPQ})
{\tilde P}^K \epsilon^{LMNPQIJ}~,
\label{eq:mshift2}\\
{\hat Q}_{IJ} &=& Q_{IJ} + 
C_{IJK}{\tilde P}^K~, \label{eq:mshift3}\\
{\hat{\tilde P}^I} &=& {\tilde P}^I~, \label{eq:mshift4}
\eea
where $I,J,\cdots = 1,\cdots,7$, and we use the notation
$Q^{IJKLM}\equiv {1\over 2}\epsilon^{IJKLMNO}Q_{NO}$.
We are interested in the special
case where the background is three intersecting $M5$-branes and, as 
explained in the end of section~\ref{scaling}, the moduli are 
restricted 
so that the ``large'' dimension $R_{7}$ does not mix with the others. 
After 
these specializations the shifts in the charges simplify 
dramatically, 
especially because there are no KK-monopoles in the background. The 
only 
nontrivial induced charges are:
\be
{\hat Z}^{i7} = {1\over 3!}C_{jkl}Q^{ijkl7}~.
\label{indm2}
\ee
The mass of the background $M5$-branes in a vacuum with general 
moduli 
can be derived by considering the effect of these induced charges, 
and 
further take a general off-diagonal metric into account. After  
computations similar to those given in Appendix~\ref{app:BPS} we
find:
\be
M^2 = Q^2_{12} + Q^2_{34}+ Q^2_{56} + ({\hat{\vec Z}})^2 + 2X~,
\label{bmass}
\ee
where:
\bea
X^2 
&=&Q^2_{56}(Q^2_{12}+({\hat Z}_{57})^{2} +({\hat Z}_{67})^{2})
+Q^2_{12}(Q^2_{34}+ ({\hat Z}_{17})^{2} +({\hat Z}_{27})^{2}) 
\nonumber \\
&~&\qquad +Q^2_{56}(Q^2_{34}+ ({\hat Z}_{37})^{2} +({\hat 
Z}_{47})^{2})
+2Q_{12}Q_{34}Q_{56}M~.
\label{xdef}
\eea
These expressions give a quartic equation for the mass which cannot
in general be further simplified. As they stand, 
(\ref{bmass}-\ref{xdef}) 
presume diagonal metric on the compact torus; however, they can be
covariantized to take off-diagonal metrics into account.

We can also construct the mass formula for the more general background
discussed in section (\ref{sec:moreg}). In this case invariants under 
duality are constructed in matrix form, simply remembering
which SL(3)'s act on the left and right.  For the background
charges, we have for the half-BPS contribution to the mass squared:
\bea
  M_0^2&=&\left[\frac{R_7(V_{135}V_{246})^{2/3}}{l_p^6}\right]^2
	\left(\tr\left[{\bf Q_o} G_o {\bf Q_o^t} \Gamma 
G_5^{-1}\Gamma\right]+
	\right.
 \nonumber \\ &~& \qquad	\left. 
 +\tr\left[{\bf Q_e} G_e {\bf Q_e^t} G_5\right]
 	+\tr\left[{\bf Q_5} G_o^{-1} {\bf Q_5^t} G_e^{-1}\right]\right)~,
\label{eq:m0}
\eea
which gives the sum of squares of the background charges
with their appropriate dependence on moduli. This expression is
covariant under motions in moduli space. Also note that the three
terms are permuted by interchange of odd and even cycles, and
by the sequence of dualities (\ref{eq:triality}).


\subsection{Fixed Scalars}
\label{fixed}
The background branes are situated in a vacuum described by $42$ 
scalar
moduli, parametrizing the coset $E_{6(6)}/USp(8)$. However, in the 
decoupling limit the value of some of these scalars are determined in 
terms of the background charges; these are the fixed scalars. The 
remaining moduli, which remain undetermined, will be referred to as 
free moduli. 
 
The fixed scalars can be characterized in general using the $N=2$ 
supersymmetry of the effective $D=5$ 
theory~\cite{Ferrara:1996um,Andrianopoli:1997pn,Andrianopoli:1997hb}. 
The general rule is that 
the scalars in the hyper-multiplets remain free moduli in the 
near-horizon 
theory, while the scalars in the vector-multiplets acquire a mass and 
become 
fixed scalars. In the toroidal compactification 
considered here there are $14$ vector-multiplets and $7$ 
hyper-multiplets. 
The vector-multiplets each have a single scalar and the
hyper-multiplets each have four scalars; so this gives $14$ fixed 
scalars and $28$ free moduli. The free moduli parametrize the
coset $F_{4(4)}/SU(2)\times USp(6)$
~\cite{Andrianopoli:1997hb,Lu:1997bg,Kutasov:1998zh}.

The physical distinction between fixed scalars and free moduli 
is exhibited clearly by minimizing the mass of the background over 
the 
full moduli space. The scalars determined this way are the fixed 
scalars; 
those that remain arbitrary are the free moduli.

\paragraph{The M5 Background:}
To see explicitly how this works, consider the $M5$-brane background.
In the simple case of a rectangular torus with
vanishing $C$-fields, the mass is:
\bea
M &=& {R_{7}\over l^6_p}(V_{12}V_{34}q_{56}
+ V_{12}V_{56}q_{34} + V_{34}V_{56}q_{12})  \\
&=& {R_{7}(V_{12}V_{34}V_{56})^{2/3}\over l^{6}_{p}}
\left( ({V_{12}V_{34}\over V_{56}^{2}})^{1/3}q_{56}+
 ({V_{12}V_{56}\over V_{34}^{2}})^{1/3}q_{34}+
  ({V_{34}V_{56}\over V_{12}^{2}})^{1/3}q_{12}\right)~,
\nonumber
\eea
where $(2\pi)^2V_{ij}$ 
is the volume of the 2-torus spanning the $(ij)$ cycle and, as before,
and $q_{ij}$ is the number of $M5$-branes wrapping the corresponding
dual cycle. The coefficient of the second equation is invariant under 
the 
dualitites of the effective $D=5$ theory, and so should not be 
varied. 
The expression in the large bracket depends only on the two 
independent 
ratios $V_{12}/V_{34}$ and $V_{34}/V_{56}$. Minimizing over these 
gives:
\bea
{V_{12}\over V_{34}} &=& {q_{12}\over q_{34}}~,
\label{vfix1} \\
{V_{34}\over V_{56}} &=& {q_{34}\over q_{56}}~.
\label{vfix2} 
\eea
A more symmetric form of the conditions is obtained by noting that the
product of the equations gives $V_{12}/V_{56} = q_{12}/q_{56}$. It is 
simple to verify these values for the fixed scalars by writing the 
explicit solutions using the harmonic function rule, and taking the 
near 
horizon limit.

We now summarize the results for more general moduli. First consider 
the 
metric, in the language of complex manifolds. The background 
$M5$-branes 
determine a holomorphic structure on the six-torus, pairing the 
indices 
$(12)$, $(34)$, and $(56)$. The components of the K\"{a}hler metric 
$G_{\mu{\bar\nu}}$ are the scalars in vector-multiplets, and 
therefore 
fixed scalars. An exception is the overall volume 
$V={\rm det}G_{\mu{\bar\nu}}$ which is a free modulus. 
(It forms a hypermultiplet together with the free modulus ${\cal 
E}_{123456}$ 
and two components of the three-form field.) The complex structure 
$G_{\mu\nu}+ {\rm h.c.}$
forms $3$ hyper-multiplets which give $12$ free moduli. Altogether 
the 
$21$ metric components therefore give $8$ fixed scalars and $13$ free 
moduli. 

Next, consider the three-form field. Expanding the mass determined
through (\ref{bmass}) for small ${\hat Z}_{i}$ gives:
\be
\delta M = {1\over 4Q} \hat{\vec{Z}}^{2}~.
\label{indmass}
\ee 
This shows that the mass has a local minimum when the induced $M2$ 
brane 
charge ${\hat Z}_{i}=0$. It is clear from (\ref{indm2}) that there 
are 
only $6$ linear combinations of $C_{ijk}$ that induce non-trivial 
$M2$-brane 
charge; this gives $6$ fixed scalar conditions, made explicit as 
follows. 
Divide the indices into three sets in accordance with the 
holomorphic 
structure, {\it i.e.} $(12)$, $(34)$, $(56)$. There are $8$ 
components of 
$C_{ijk}$ that have one index in each set; all these are free 
moduli. 
The remaining 12 components each have two indices within one of the 
sets, 
and the remaining index in a different set, {\it e.g.} $C_{125}$. 
The $6$ fields of this kind that are selfdual on the $T^4$ transverse 
to 
the unpaired index are fixed scalars, their anti-selfdual partners 
are 
free moduli. For example, 
$C_{{\hat 1}{\hat 2}5}+C_{{\hat 3}{\hat 4}5}$ 
is a fixed scalar, but 
$C_{{\hat 1}{\hat 2}5}-C_{{\hat 3}{\hat 4}5}$ is a free modulus. 

In summary, the $14$ fixed scalars are $8$ components of the
metric and $6$ components of the three-form field; and the
$28$ free moduli are $13$ components of the metric, $14$ components
of the three-form, and the $1$ pseudoscalar ${\cal E}_{123456}$.

Another way of stating the fixed scalar conditions is that 
the three physical charges are identical:
\be
Q_{12}=Q_{34}=Q_{56}\equiv Q
={R_{11}\over l^6_P} V^{2\over 
3}~(q_{12}q_{34}q_{56})^{1\over 3}~,
\ee
where the volume $V=V_{12}V_{34}V_{56}$. The mass of the background
at the fixed scalar point is $M_{\rm fix} = 3Q$. 

\paragraph{The Type IIB Background:}
The distinction between fixed scalars and free moduli is in some ways 
simpler 
in the type IIB F1/NS5/KK-monopole background, so we sketch the 
results
in this case too. On a rectangular torus simple minimization, as 
above, gives two obvious fixed scalars:
\bea
r^2_5 &=& {q_5\over q_k}~, \label{eq:fixr5} \\
{v_4\over g^2_{s}} &=& {q_1\over q_5}~,\label{eq:fixg6}
\eea
where $r_{5}$ the radius of the KK-monopole direction in string units,
$v_{4}$ is the volume of the internal $T^{4}$ also in string units,
and $g_{s}$ is the string coupling constant. These fixed scalar 
conditions 
agree with the result of simply reading off the near-horizon values 
of the
explicit metric, written using the harmonic function rule.

More generally, consider the $26$ NS scalars 
$e^{\Phi_{6}}$ and $G_{ij}+B_{ij}$ (with 
$i,j=1,\cdots 5$). Of these, $20$ are free moduli, namely 
$G_{ij}+B_{ij}$ 
with $i=1,\cdots,4$ and $j=1,\cdots, 5$; these parametrize the coset 
$SO(4,5)/SO(4)\times SO(5)$. The remaining $6$ NS-moduli are fixed 
scalars.
On a rectangular torus this was shown explicitly for $G_{55}$ and 
$e^{\Phi_{6}}$ above, and for $G_{i5}-B_{i5}$ 
in~\cite{Kutasov:1998zh}.

The $16$ RR scalars can be represented as a formal sum of forms
$\chi + C^{(2)} + A$. The anti-selfdual part of this form are
$8$ fixed scalars, and the selfdual part gives $8$ free moduli. 
In components, the fixed scalars are $C^{-}_{ij}$ and $\chi-A_{1234}$
and $C_{i5}-{1\over3!}\epsilon_{ijkl}A^{jkl}_{~~~5}$ with 
$i,j\cdots=1,\cdots,4$; and the free RR-moduli are given by similar 
expressions, but with relative signs flipped.

Altogether there are $6+8=14$ fixed scalars and $20+8=28$ free 
moduli, as there should be.

\paragraph{More General Backgrounds:} 
As a final example of the determination of fixed scalars, we consider
the more general backgrounds discussed in section (\ref{sec:moreg}).
For a given choice of $9$ charges there are two sets of moduli
that are relevant; for example, for the pure M5 configurations
represented by the matrix ${\bf Q_5}$, both the odd and the even 
metric couple, 
for a total of $10$ moduli; or for the backgrounds
involving ${\bf Q_o}$, one needs $G_o$ and $G_5$.
  
The fixed scalars are determined by extremizing the mass formula for 
the 
background. In the present case there are no parity-odd moduli so the
induced charges (\ref{indm2}) vanish. The mass formula becomes:
\be
  M^2 = M^{2}_{0} + 2X~,
\label{genmass}
\ee
where $M^{2}_{0}$ was given in (\ref{eq:m0}) and:
\bea
X^2&=&
  \Bigl(\tr[{\bf Q_5}G_o^{-1}{\bf Q_5^t} G_e^{-1}]\Bigr)^2
	-\tr[({\bf Q_5}G_o^{-1}{\bf Q_5^t}G_e^{-1})^2]
	+2 {\rm det}({\bf Q_{5}})M~.
\label{xdefgen}
\eea
The determinant of ${\bf Q_{5}}$ does not depend on the moduli,
and both $M_{0}^{2}$ and $X^{2}$ depend only on the combination 
$Y={\bf Q_5}G_o^{-1}{\bf Q_5^t}G_e^{-1}$. The mass $M$ therefore
depends only on $Y$, as far as its dependence on moduli is 
concerned.  Now, one can check that the trace of any power of the 
quantity $\tr[Y^n]$ is extremized by:  
\be
  G_o={\bf {\hat Q}_5}^tG_e^{-1}{\bf {\hat Q}_5}\quad,\qquad 
	{\bf {\hat Q}_5}={\bf Q_5}/\det[{\bf Q_5}]^{1/3}~.
\label{fxsclreq}
\ee
It follows that, whatever the exact expression for the
mass in terms of $Y$, its extremum is \pref{fxsclreq}. This shows
that, of the two $SL(3)/SO(3)$ cosets that are ``active'' in the
$9$ charge background, the fixed scalar equations determine one
in terms of the other. In the subspace of moduli space considered here
there are thus five free moduli, and we are able to find the relation 
to the other five active moduli explicitly.


\subsection{Instability under Fragmentation}
As discussed in section~\ref{sec:masssec}, the mass depends 
nonlinearly 
on the charges at generic points in moduli space. The background is 
only unstable under fragmentation when its mass is equal to the sum 
of its 
constituent masses so this effect is sufficient to stabilize the 
configuration. The surface of instability is therefore some 
lower-dimensional surface in moduli space~\cite{Maldacena:1999uz}. 

\paragraph{The M5-Background:} To characterize this surface 
explicitly,
consider (\ref{indmass}) for the mass due to induced charges. 
In the ground state of the background the fixed scalar conditions 
tune 
these charges to zero, {\it e.g} 
${\hat Z}^{1}=C^{134}Q_{34}+C^{156}Q_{56}=0$.
However, the ratio of charges $Q_{34}/Q_{56}$ is generally different
for each of the fragments in the final state, so the decay products 
are 
heavier than they would be for vanishing three-form field. This is 
the 
effect that stabilizes the background configuration at generic points 
in
moduli space.  In order for fragmentation to be allowed,
there are $6$ conditions of this sort on the three-form 
field, and $6$ additional conditions that arise similarly from the 
fixed scalar equations for the off-diagonal metric. 
More precisely there are three hyper-multiplets (with four scalars 
apiece) that each protect against emission of two kinds of charges, 
but 
not the third. If any two of the hyper-multiplets vanish the 
configuration 
is unstable. The surface of instability is therefore a codimension $8$
subspace of moduli space.

It follows from this reasoning that the background is unstable 
{\it everywhere} in moduli space, when all the decay products have 
the same ratios of charges as the initial state. We avoid this 
degenerate
possibility throughout this work, by assuming that the three 
background 
charges are mutually prime.

\paragraph{The Type IIB Background:}
It is simpler to characterize the surface of instability
in the canonical type IIB configuration F1/NS5/KK-monopole. Here it 
is 
manifest that there is a linearly
realized $SO(5,4)$ symmetry acting on the $20$ free NS-moduli 
$G_{ij}+B_{ij}$ (with $i=1,\cdots,5$, $j=1,\cdots,4$). These moduli 
do 
not affect the linear dependence of the mass formula on the charges,
so turning on these moduli does not prevent fragmentation.
The RR-moduli are quite different: they enter nonlinearly in the way
discussed explicitly above for the canonical $M5$-brane 
configuration. 
One component of the co-dimension $8$ surface of instability is 
therefore 
precisely the surface where the $8$ free RR-moduli vanish. It is 
parametrized by the $20$ free NS-moduli.

\subsection{The Global Structure of Moduli Space}
\label{glob}
The $28$ free moduli parametrize a moduli space which is {\it 
locally} the
non-compact coset $F_{4(4)}/SU(2)\times USp(6)$ .
The purpose of this subsection is to discuss 
the {\it global} structure of this space.
Naively, one might expect to be able to quotient 
by $F_{4(4)}(\IZ)$ but, as explained in~\cite{Seiberg:1999xz,lm99}
for the D1-D5 case, this group is too large -- it does not leave
the set of background masses invariant.

An explicit prescription of the identifications for the
D1-D5 system was given in~\cite{lm99}.
There, the local structure of the moduli space
is $SO(5,4)/SO(5)\times SO(4)$. It is sufficient to
consider a representative $SL(2,\IZ)\times SL(2,\IZ)$ 
subgroup of the $SO(5,4,\IZ)$ duality group. Under this subgroup,
a given pair of D1-D5 charges $(q_1,q_5)$ can be mapped
to a ``canonical background'' $(N=q_1q_5,1)$.
The global identifications of the moduli space are
those that preserve the charge vector of the canonical
background.  This subgroup of $SO(5,4;\IZ)$ is generated by
a particular diagonal $\Gamma_0(N)$ subgroup of $SL(2,\IZ)\times 
SL(2,\IZ)$,
together with conjugations by elements of the T-duality group.
The interesting part of the structure
can be projected onto the $SL(2,\IR)/SO(2)$ subspace of the
moduli space acted on by one of the $SL(2,\IZ)$'s;
the fundamental domain projected onto this subspace is
then $\Gamma_0(N)\backslash SL(2,\IR)/SO(2)$.
The moduli space has a cusp for each factorization of
$N$ into one-brane and five-brane charges $q_1$, $q_5=N/q_1$,
where there is a weakly-coupled (large target space)
sigma model description of the dynamics.
The singular locus, where the D1-D5 bound state can fragment
into its constituents, is a codimension four subspace
of the moduli space where the sigma model description
breaks down (even at large volume) due to singularities
in its target space.  Under the projection, the singular
loci consist of geodesics running between conjugate
cusps for D1-D5 charges $(q,N/q)$ and $(N/q,q)$.

One can employ the same strategy 
in the present context; now the $SL(3,\IZ)^{3}$ subgroup
of $E_{6(6)}(\IZ)$ duality is representative.  
Many features carry through.
First of all, consider a fivebrane charge matrix~\pref{fivecharge}:
\be
  {\bf Q_5}=\pmatrix{q_{12} & 0 & 0 \cr
	   0 & q_{34} & 0 \cr
	   0 & 0 & q_{56} }\ .
\ee
This can be mapped to a ``canonical background''
${\bf Q_5}={\rm diag}[N=q_{12}q_{34}q_{56},1,1]$ via the
$SL(3,\IZ)\times SL(3,\IZ)$ transformation 
${\bf Q_5}\rightarrow g_L {\bf Q_5} g_R^t$,
with:
\bea
  g_L&=&\pmatrix{a'q_{34}q_{56}& -ab'q_{12}q_{56} & bb'q_{12}q_{34} 
\cr
		c'	& ad'q_{56}	& -bd'q_{34}   \cr
		0	& c		& d	    }~,
  \nonumber\\
& &\nonumber\\
  g_R&=&\pmatrix{d'q_{34}q_{56}& -dc'q_{12}q_{56}& cc'q_{12}q_{34} \cr
		b'	& da'q_{56}	& -ca'q_{34}   \cr
		0	& b		& a	    }~,
\label{canontransf}
\eea
where the coefficients satisfy:
\be
 adq_{56}+bcq_{34}=a'd'q_{34}q_{56}+b'c'q_{12}=1~.
\ee
For the existence and uniqueness of this transformation
for any $(q_{12},q_{34},q_{56})$ with $q_{12}q_{34}q_{56}=N$,
we require that the prime decomposition of $N$
contain any given prime no more than once.
Without loss of generality, one can set \eg\ $a=b=a'=b'=1$.
The canonical charge matrix is preserved by further
transformations generated by a copy of $\Gamma_0(N)$:
\bea
  {\hat g}_L&=&\pmatrix{
	\alpha &	-\beta q_{12}q_{34}q_{56} &	0 \cr
	\gamma &	\delta	& 0	\cr
	0	& 0	& 1 }
  \quad,\qquad \alpha\delta+\beta\gamma q_{12}q_{34}q_{56}=1 
\nonumber\\
\nonumber\\
  {\hat g}_R&=&\pmatrix{
	\delta &	-\gamma q_{12}q_{34}q_{56} &	0 \cr
	\beta &	\alpha	& 0	\cr
	0	& 0	& 1 }\quad ,
\label{furthertransf}
\eea
together with the $SL(2,\IZ)$ that acts in the lower right corner.
The fixed scalar conditions identify the moduli subspaces
acted on by the left and right $SL(3,\IZ)$ duality subgroups;
the identifications of this five-dimensional space 
under the action of $\hat g$ is representative of the
global structure of the moduli space, just as in
the D1-D5 case.  The full residual duality 
group $\HH_N\subset F_{4(4)}(\IZ)$ is generated (in the IIB
frame where the background charges are purely NS)
by the subgroup of $SL(3,\IZ)$ analogous to the above, 
together with T-duality transformations $SO(5,4;\IZ)$.

The singular locus in this subspace are
the diagonal matrices in $SL(3,\IR)/SO(3)$,
since then there is no projection of any given fivebrane
charge onto the worldvolume of another.  From
the $A\cdot N$ decomposition, one sees that this is a two-dimensional
geodesic submanifold of the five-dimensional homogeneous space.
There will be such a submanifold for any unordered triple
of fivebrane charges $(q_{12},q_{34},q_{56})$, which gets
mapped to a unique geodesic submanifold under
the transformation~\pref{canontransf} to the
canonical background.  The generic point on
the boundary boundary of $SL(3,\IR)/SO(3)$
is reached when $A={\rm diag}[\lambda_1,\lambda_2,\lambda_3]$
(with $\lambda_1\lambda_2\lambda_3=1$)
has one of its eigenvalues degenerate, say 
$\lambda_1\rightarrow\infty$.
Clearly this can happen to any one of the three $\lambda_i$,
with the possibilities permuted by the action of the Weyl
group in $SL(3)$; this is why the singular
loci correspond to unordered triples of constituent charges,
since each of these degenerations belongs to the same
geodesic submanifold but different orderings of the $Q_i$.

The singular loci corresponding to different unordered triples
$(q_{12},q_{34},q_{56})$ do not intersect, by the same
argument given in~\cite{lm99} for the D1-D5 system.
Take the background charges to be those of 
F1/NS5/KK-monopoles
in the type IIB description.
In this case, the singular locus corresponds to vanishing
of all RR moduli, and is left invariant only
by T-duality transformations $SO(5,4;\IZ)$.  
Assume that there is a nontrivial intersection, 
in the canonical background, of the codimension eight
singular subspaces corresponding to two different unordered
triples of background charges.
Pull a point on the intersection back to the original background
by the maps specified by one of the two charge sets.
The resulting point is parametrized only by NS moduli;
but the charges $(q_{12},q_{34},q_{56})$ and 
$(q_{12}',q_{34}',q_{56}')$ 
cannot be related by T-duality, so we conclude that the
singular loci corresponding to different unordered
triples of charges are disjoint.


\section{The Charged Excitations}
\label{exc}
In any background preserving 1/8 of SUSY there is a rich variety of 
charged 
excitations, as discussed in section~\ref{scaling}. The exploration
of this spectrum is one way to learn about 
the structure of the theory governing the background.

\subsection{Introduction to the Spectrum of Charges} 
\label{tempcur}
One of the most basic properties of the charged excitations is their
spectrum, {\it i.e.} the energy as a function of the charge vector.
In this section we present heuristic arguments that motivate the 
spectrum in the case of a rectangular torus; a more precise 
computation 
is given in Appendix~\ref{app:BPS}. 

Consider the type IIB F1/NS5/KK-monopole background and include 
momentum
along the fundamental string. The mass of this system is simply the 
sum 
of constituent masses:
\be
M = Q_K + Q_5 + |\vec{Q}_1 + \vec{P}_{\rm tot}|~.
\ee
In this formula we have exploited the rotational invariance of the 
NS5/KK-monopole world-volume to allow a general direction of the 
vector $\vec{Q}_1 + \vec{P}_{\rm tot}$. In the scaling limit the 
component 
of this  vector along the "large" dimension $R_6$ is dominant; it 
corresponds to the background fundamental string with charge $Q_{1}$,
and a neutral chiral excitation.
We denote the remaining four components $\vec{W}_{F1} + \vec{P}$; 
these 
correspond to charged excitations. After expansion, the
energy of the charged excitations becomes:
\be
E = {1\over 2Q_1} ( \vec{W}_{F1} + \vec{P})^2~.
\label{eq:spec1}
\ee
An important qualitative consequence of this result is that the 
charged excitations have {\it finite} energy in the environment 
created 
by the background, even though their masses in flat ambient space 
{\it diverge} 
in the scaling limit. It is shown in Appendix~\ref{app:BPS} that the 
expression 
(\ref{eq:spec1}) remains valid for general orientations of the charge 
vectors $\vec{W}_{F1},\vec{P}$.

The duality $ST_{1234}S$ leaves the background invariant, except
for the interchange of two charges $Q_1\leftrightarrow Q_5$. Thus
(\ref{eq:spec1}) implies that the theory has excitations with 
the spectrum:
\be
E = {1\over 2Q_5} ( \vec{W}_{D1} + \vec{W}_{D3})^2~,
\label{eq:spec2}
\ee
where $\vec{W}_{D1}$ is the charge vector of $D1$-branes wrapping 
$1234$; and the $\vec{W}_{D3}$ denote three-branes wrapped within
$1234$, with the vector index being the direction that is {\it not} 
wrapped. The further duality $T_{15}$ similarly leads to the 
spectrum:
\be 
E = {1\over 2Q_K} ( \vec{W}_{D} + \vec{W}_{\tilde D})^2~,
\label{eq:spec3}
\ee
where the charge vectors are:
\bea
\vec{W}_{D}&=&(D1_5,D3_{125}, D3_{135},D3_{145})~,\\
\vec{W}_{\tilde D}&=&(D5_{12345},D3_{345}, D3_{245},D3_{235})~.
\eea

Recall from section \ref{scaling} that the total number of $U(1)$ 
charges 
is $27$. The formulae above gives the energy of $24$ of these. The 
remaining three charges are more complicated, because they are the
electric-magnetic duals of the charges that appear in the background.
The spectrum of these ``special'' charges:
\be
E^{\rm spec}= {1\over 2(Q_1 + Q_5+ Q_K)} (F_5 + P_5+ N_{12345})^2~,
\ee
is derived in Appendix~\ref{app:BPS}. It is also shown that the 
energy 
of a configuration with general charge vector is the sum of the 
special cases considered above. The final result for the energy of
excitations with general charge vector therefore becomes:
\bea
E &=& {1\over 2Q_1} ( \vec{W}_{F1} + \vec{P})^2
+ {1\over 2Q_5} ( \vec{W}_{D1} + \vec{W}_{D3})^2
+ {1\over 2Q_K} ( \vec{W}_{D} + \vec{\tilde W}_{D})^2 \nonumber \\
&~&\qquad + {1\over 2(Q_1 + Q_5+ Q_K)} (F_5 + P_5+ N_{12345})^2~.
\label{eq:genenergy}
\eea

\paragraph{Microscopic Units:}
So far we have used physical units where the charge is identical to 
the 
mass of an isolated brane and is denoted by a capital letter. The 
microscopic units instead count the number of branes and are denoted 
by 
lower-case letters. In these units the energy (\ref{eq:genenergy}) 
becomes:
\bea
R_6 E &=& {1\over 2q_1} ( w^{F1}_ i   r_i+ p^i /r^i)^2
+ {1\over 2q_5} ( w^{D1}_i   {r^i\over\sqrt{v_4}}+ 
\vec{w}_{D3}{\sqrt{v_4}\over r^i})^2 
+ {1\over 2q_k} ( w^{D}_i e^i+ w^{i}_{\tilde D} /e^i)^2 \nonumber \\
&~&\qquad + {1\over 6q_1 q_5 q_k} ( f_{5}q_{5} + p^{5}q_{k} +
n_{12345}q_{1})^2~,
\label{eq:micspec}
\eea
where, in each of the first terms, a sum over the index $i$ is 
implied {\it after} taking the square. 
The last term (containing the special $U(1)$ charges)
was rewritten using the fixed scalar equations 
(\ref{eq:fixr5}-\ref{eq:fixg6}).
The $r^i$ are radii of the 
compact dimensions in string units ({\it i.e.} $r^i=R_i/l_s$ where 
$l_s=\sqrt{\alpha^\prime}$), $v_4\equiv r_1 r_2 r_3 r_4$, and 
$e^i = {1\over\sqrt{v_4}}(1,r_1r_2,r_1r_3,r_1r_4)$. It is important
to note that the scale of the energy is set by the radius of the large
dimension $R_6$.

\paragraph{The M5-Background:}
It is straightforward to find the corresponding formulae for the 
three M5-branes intersecting over a line, {\it e.g.} using the duality
(\ref{eq:duality}). The result is:
\bea
E  &=& {1\over 2Q_{56}}\left[(Z_{13}+Z_{24})^2
+(Z_{14}+Z_{32})^2 + (P_5 + Q_{67})^2+ (P_6 + 
Q_{57})^2\right] 
\nonumber \\
&~&~~+ {1\over 2Q_{12}}\left[(Z_{35}+Z_{46})^2
+(Z_{36}+Z_{54})^2 + (P_1 + Q_{27})^2+ (P_2 + Q_{17})^2\right] 
\nonumber \\
&~&~~+ {1\over 2Q_{34}}\left[(Z_{15}+Z_{26})^2
+(Z_{16}+Z_{25})^2 + (P_3 + Q_{47})^2+ (P_4 + Q_{37})^2\right]
\nonumber \\
&~&~~+ {1\over 2(Q_{12}+Q_{34}+Q_{56})}
(Z_{12}+Z_{34}+Z_{56})^2~,
\eea
where $Z_{ij}$ are the charges of the M2-branes wrapping the $(ij)$ 
cycle,
$P_i$ is the momentum along $R_i$, and $Q_{ij}$ denote five-branes 
transverse to the $(ij)$ cycle. The intersecting $M5$ background 
makes 
the systematics of the energy formula clearer:
the $3$ special $U(1)$ charges are the $M2$-branes that are
``electric'' duals of the ``magnetic'' background $M5$'s. 
The $8$ $U(1)$ charges that are weighted by a given background $M5$
($Q_{12}$, $Q_{34}$, or $Q_{56}$) are the $M2$'s that do not share an
index with the background $M5$; and $KK$-excitations and light $M5$'s 
that do. These statements translate into simple geometric relations.

\subsection{Currents and Lattices}
\label{currents}
The purpose of this section is to take a first step towards an 
interpretation of the charged spectrum (\ref{eq:micspec}). The 
working 
assumption is that the spectrum arises as the 0-modes of affine
currents in the underlying conformal field theory. The AdS/CFT 
correspondence~\cite{adsrev}, as implemented for 
$AdS_{3}$ in~\cite{Giveon:1998ns,deBoer:1998pp,Kutasov:1999xu}, 
shows that this is 
the correct interpretation for the perturbative 
currents; non-perturbative dualities then 
suggest that all currents arise in this way.

Consider the first three terms of (\ref{eq:micspec}). These terms 
have 
the structure of $12$ $U(1)$ currents in the right-moving sector, 
$4$ currents at each level $q_{1}$, $q_{5}$, and $q_{k}$. Modular 
invariance of the CFT then requires a matching set of $12$ 
left-moving currents with spectrum:
\be
R_6 E_{L} = {1\over 2q_1} ( w^{F1}_ i   r_i - p^i /r^i)^2
+ {1\over 2q_5} ( w^{D1}_i   {r^i\over\sqrt{v_4}} - 
\vec{w}_{D3}{\sqrt{v_4}\over r^i})^2
+ {1\over 2q_k} ( w^{D}_i e^i - w^{i}_{\tilde D} /e^i)^2~.
\label{eq:eltemp}
\ee
Taken together, the right- and left-moving currents form a simple 
lattice of signature $(12,12)$. The right-moving currents
are invariant under the duality $T_{1234}$; the left-moving ones
change sign. It is important to note that these results are reliable 
only
in the semiclassical regime: there is no supersymmetry in the 
left-moving 
sector, so the formula (\ref{eq:eltemp}) can not be interpreted as a 
BPS 
formula. It can therefore receive corrections; in particular, the 
levels may 
receive contributions at the subleading order in the background 
charge. 

Next, we turn attention to the three ``special'' charges, appearing 
in the last line of (\ref{eq:micspec}). An important constraint on 
these
is that the CFT has a global $SU(2)\times USp(6)$ 
symmetry~\cite{Larsen:1998xm}. The full set of charges transforms
in the ${\bf 27}$ of $USp(8)$ which decomposes as 
$({\bf 2},{\bf 6})\otimes ({\bf 1},{\bf 14})\otimes ({\bf 1},{\bf 
1})$ 
under $SU(2)\times USp(6)$ . The global $SU(2)$ acts on the 
supersymmetries, 
so the $({\bf 2},{\bf 6})$ is identified with 
the $12$ right-moving charges. More generally, the energy formula 
(\ref{eq:micspec}) is the extremality condition on the right-moving 
charges. 
This identifies a specific linear combination of the special charges 
as 
right-moving; and we further note that this combination is the
singlet $({\bf 1},{\bf 1})$ under $SU(2)\times USp(6)$. 
At this point we have yet to account for the $12$ left-moving currents
with spectrum given in (\ref{eq:eltemp}),  and two linear combinations
of the special charges. These must transform as $({\bf 1},{\bf 14})$ 
under the
$SU(2)\times USp(6)$ global symmetry and this property determines the
spectrum of the special charges as:
\be
R_{6} E_{L}^{\rm spec} =
{1\over 2q_1 q_{5} q_{k}} ( f_5  q_{5} - p_{5} q_{k})^2+
{1\over 6q_1 q_{5} q_{k}} ( f_5  q_{5} + p_{5} q_{k}-2 
n_{12345}q_{1})^2~.
\label{eq:elspec}
\ee
The first term in this equation is the dual of the first term in 
(\ref{eq:eltemp}), after the fixed scalar condition (\ref{eq:fixr5}) 
is 
taken into account. This expresses the perturbative $SO(5,4)$ 
duality. The second term is determined as the linear combination of 
charges that are orthogonal both to the perturbative charges in the 
first term, and the $USp(6)$ singlet; it must be normalized as the
other elements in the ${\bf 14}$ of $USp(6)$. As a further check on 
(\ref{eq:elspec}), note that the total left moving weight is 
symmetric 
under triality of the charges, an obvious symmetry in the 
M-representation. 

In addition to the $27$ $U(1)$ charges there is the bulk angular 
momentum of the dual black hole in five dimensions. This charge is 
interpreted in the CFT as a component of the $SU(2)$ R-charge, with 
normalization fixed by the supersymmetry 
algebra~\cite{Breckenridge:1996is,Horowitz:1996ac,Dijkgraaf:1997it}. 
The total spectrum of the ``special'' 
right-moving charges therefore becomes:
\be
R_{6} E_{R}^{\rm spec} = {1\over 2q_1 q_{5} q_{k}} J^{2}+
{1\over 6q_1 q_{5} q_{k}} ( f_5  q_{5} + p_{5} q_{k}+ 
n_{12345}q_{1})^2~.
\label{eq:erspec}
\ee
There is an obvious similarity between (\ref{eq:elspec}) and 
(\ref{eq:erspec}).
This suggests that the special currents are incorporated into
the $(12,12)$ lattice found previously so, altogether, the currents
form a $(14,14)$ lattice.

There are two peculiar features of the spectrum of excitations
(\ref{eq:erspec}) carrying the special charge:
First of all, the normalization differs by a factor of two relative to
the other currents (in particular the last term in \pref{eq:elspec});
secondly, although this special charge appears in the 
right-moving energy, it is associated to the graviphoton
tower of supergravity on $AdS_3\times S^2$, which has
the opposite chirality (it is left-moving)%
\footnote{The latter fact appears to violate the rule of thumb 
that the singleton sector can be obtained 
by extending the range of the mode index in the KK tower.}
~\cite{Larsen:1998xm,deBoer:1998ip}.
Nevertheless, these results are dictated by BPS algebra and
global symmetries.

\subsection{Black Hole Entropy} 
\label{bhcft}
The excitations that we consider correspond at strong coupling
to regular black holes. It is known that the microscopic 
(statistical) 
entropy of these black holes agrees with the Bekenstein-Hawking area 
formula, when only the scalar charge (momentum) is excited. It is 
interesting to generalize this agreement to the case where the 
additional 
$U(1)$ charges are included as well. We carry out the computation in 
the 
type IIB background, on a rectangular torus. Some elements of the more
general computation are discussed in Appendix~\ref{sec:sl3}, 
and others are given 
in~\cite{lm99}.

\paragraph{Conformal Field Theory:}
For a general background the central charge of the underlying
CFT is proportional to the unique cubic invariant of $E_{6(6)}$ 
that can be formed from the charge vectors in the fundamental 
representation ${\bf 27}$. For the present purposes it is sufficient
to consider the canonical F1/NS5/KK-monopole background;
then the central charge is simply $c = 6 q_1 q_5 q_k$. 

The operators of the theory have left and right conformal 
weights $h_{L,R}={1\over 2}(\epsilon\pm p_6)$ where $\epsilon$ is the 
total 
energy. In a sector with a specified $U(1)$ charge vector
the typical vertex operator can be written:
\be
V_{\rm tot} = V_{\rm irr}~V_{\rm U(1)}~,
\ee
where $V_{\rm U(1)}$ carries the required $U(1)$ charge and
$V_{\rm irr}$ is neutral, but otherwise arbitrary. This shows how
a part of the conformal weight is expended on exciting the $U(1)$ 
charges.
The unspecified neutral excitations provide the microscopic 
degeneracy.
The corresponding ``irreducible'' conformal weights are:
\bea
h^{\rm irr}_L &=& {\epsilon+p_6\over 2}-
{1\over 4q_1}(\vec{p}-\vec{w}_{F1})^2 -
{1\over 4q_5}(\vec{w}_{D3}-\vec{w}_{D1})^2-
{1\over 4q_k}(\vec{w}_{D}-\vec{w}_{\tilde D})^2 \nonumber \\
&~&\qquad
-{1\over 4q_1 q_5 q_k}(f_5 q_{5}-p_{5}q_{k} )^2 
-{1\over 12q_1 q_5 q_k}( f_5 q_{5} + p_{5}q_{k} -2n_{12345}q_{1})^2 
\label{hl} \\
h^{\rm irr}_R &=&  {\epsilon-p_6\over 2}- 
{1\over 4q_1}(\vec{p}+\vec{w}_{F1})^2 -
{1\over 4q_5}(\vec{w}_{D3}+\vec{w}_{D1})^2-
{1\over 4q_k}(\vec{w}_{D}+\vec{w}_{\tilde D})^2 \nonumber \\
&~& \qquad -{1\over 12q_1 q_5 q_k}( f_{5}q_{5}+ p_{5}q_{k}
+ n_{12345}q_{1})^2  -{1\over 4q_1 q_5 q_k}J^2
\label{hr} 
\eea
where contractions of vectors employ the appropriate moduli-dependent 
metric, given in section~\ref{tempcur}. The entropy is given in terms 
of 
the conformal weights as:
\be
S = 2\pi\left[ \sqrt{ch^{\rm irr}_L\over 6} +
\sqrt{ch^{\rm irr}_R\over 6}\,\right]~.
\label{eq:gmicro}
\ee
This is a fairly intricate function of the various black hole 
parameters.

In the extremal limit the energy $\epsilon$ is determined
such that $h^{\rm irr}_R=0$ so the entropy becomes:
\be
S = 2\pi\sqrt{q_{1}q_{5}q_{k}h^{\rm irr}_{\rm ext}}~.
\label{eq:microext}
\ee
The left conformal weight can be written:
\bea
h^{\rm irr}_{ext} &=& p_6 + {1\over q_1}\vec{p}\cdot\vec{w}_{F1}
+ {1\over q_5}\vec{w}_{D3}\cdot\vec{w}_{D1}
+ {1\over q_k}\vec{w}_{D}\cdot\vec{w}_{\tilde D} +
{1\over 4q_{1}q_{5}q_{k}}J^{2} \\
&~&~ - f^2_5 {q_5\over 4q_k q_1}- p^2_5 {q_k\over 4q_1 q_5}- 
n^2_{12345} {q_1\over 4q_k q_5}
+ {1\over 2q_1}f_5 p_5+ {1\over 2q_k}f_5 n_{12345}+ 
{1\over 2q_5}p_5 n_{12345}\nonumber 
\eea
Note that the signs are such that the last terms in this expression
do {\it not} form a complete square. 

\paragraph{The Area Formula:}
The entropy formula should be compared with the entropy that follows 
from the area of the corresponding macroscopic black holes. However,
the classical solutions needed are difficult to construct, and the
task has not been completed. The difficult features are those related
to the special charges, {\it i.e.} the magnetic duals of the 
background
charges. In the nonrotating case, a generating solution has been 
constructed which in principle contains the required 
data~\cite{Cvetic:1995kv}.
However, it is given in a form which makes the area formula hard to 
disentangle; for discussion and further references 
see~\cite{Cvetic:1997im}.
In view of these problems we restrict ourselves to the BPS-limit 
$h^{\rm irr}_R=J=0$ where the black hole entropy is known 
on general grounds. It is~\cite{Kallosh:1996uy}:
\be
S = \pi\sqrt{J_4}~,
\label{eq:area}
\ee
where $J_4$ is the quartic invariant of $E_{7(7)}$:
\be
-J_4=
x^{ij}y_{jk}x^{kl}y_{li}-{1\over 4}x^{ij}y_{ij}x^{kl}y_{kl}
+{1\over 96}\epsilon_{ijklmnop}(x^{ij}x^{kl}x^{mn}x^{op}
+y^{ij}y^{kl}y^{mn}y^{op})~.
\label{eq:j4}
\ee
The invariant is written in the $SO(8)$ formalism where the central 
charges can be immediately identified with wrapped 
branes~\cite{Balasubramanian:1998az}. In M-theory notation the map is 
$x^{ab}=z^{ab}$, $x^{a8}={\tilde p}^{a}$, $y_{ab}=q_{ab}$, 
$y_{a8}=p_a$ where the indices $a,b,\cdots = 1,\cdots,7$. 
After specializing to the background with three $M5$'s intersecting 
over a line, with further $U(1)$ charges turned on, this 
becomes\footnote{The signs are not strictly in accord with 
(\ref{eq:j4}). 
The convention in this paper is to assign brane (vs. anti-brane) 
numbers 
so that a maximal number of terms contribute positively to the 
entropy; 
this simplifies dualities.}:
\bea
J_4 &=& 4q_{12}q_{34}q_{56}p_{7}\nonumber \\
 &~& +  4q_{34}q_{56}\left[q_{57}p_6+q_{67}p_5
+z^{13}z^{24}+z^{14}z^{23}\right]\nonumber \\
&~& +  4q_{34}q_{56}\left[q_{17}p_2+q_{27}p_1
+z^{35}z^{46}+z^{36}z^{45}\right]
\nonumber \\
 &~& +  4q_{12}q_{56}\left[q_{37}p_4+q_{47}p_3
+z^{15}z^{26}+z^{16}z^{25}\right]\nonumber \\
&~&- (q_{12}z^{12})^{2}
- (q_{34}z^{34})^{2}- (q_{56}z^{56})^{2}\nonumber \\
&~&+
2q_{12}z^{12}q_{34}z^{34}
+2q_{12}z^{12}q_{56}z^{56}
+2q_{56}z^{56}q_{34}z^{34}~.
\label{eq:redj4}
\eea
The area formula (\ref{eq:area}), with the reduced quartic invariant 
(\ref{eq:redj4}), is identical to the microscopic expression
(\ref{eq:microext}) after dualities. This 
agreement gives some confidence that the conformal weights have
been correctly identified also in the non-BPS case.

\subsection{Parity-odd Moduli} 
The discussion of the spectrum of charged excitations so far assumed
a choice of moduli corresponding to a rectangular torus. It is 
straightforward to generalize and take into account off-diagonal 
metric
components on the compact space: simply contract indices using the 
general metric, subject to the fixed scalar conditions.

The parity-odd moduli are more interesting. They induce
shifts in the $U(1)$ charges that can be computed using the general
rules given in (\ref{eq:mshift1}-\ref{eq:mshift4}). After 
specializing to the moduli that respect the effective five-dimensional
structure these formulae show that the background charges and the 
charged 
excitations do not mix. The charged excitations shift via: 
\bea
{\hat P}_i &=& P_i + {1\over 2}C_{jki}Z^{jk}
+ ({1\over 4!}C_{jkl}C_{mni}+{1\over 5!}{\cal E}_{jklmni})
Q^{jklmn}~,
\label{eshift1}\\
{\hat Z}^{ij} &=& Z^{ij} + {1\over 3!}C_{klm} Q^{klmij}~,
\label{eshift2}\\
{\hat Q}^{ijklm} &=& Q^{ijklm}~.
\label{eshift3}
\eea
The shifts in the background charges are compensated by changes in 
the 
fixed scalars. The net result is therefore that the spectrum of 
the charged excitations is modified by the parity-odd moduli through
the shifts (\ref{eshift1}-\ref{eshift3}) of the $U(1)$ charges, but 
with the background charges unmodified. This rule provides rather
detailed information about the structure of the spacetime CFT.

\subsection{General Moduli and the Charged Excitations}
\label{sec:sl3}
The moduli dependence of the excitations can be investigated in more
detail using the setup introduced in section~\ref{sec:moreg}. The 
discussion of the background given there is extended to the 
excitations by introducing the matrix expressions:
\be\label{oddex}
 {\bf Z_o}=\pmatrix{
 p_{1} & p_{3} & p_{5} \cr
 z_{35} & z_{15} & z_{13} \cr
 q_{17} & q_{37} & q_{57}} \ ,
\ee
\be\label{evenex}
 {\bf Z_e}=\pmatrix{
 {p}_2 & {p}_4 & {p}_6 \cr
 z_{46} & z_{26} & z_{24} \cr
 q_{27} & q_{47} & q_{67}} \ ,
\ee
as well as:
\be\label{twoex}
{\bf Z_2}=\pmatrix{z_{12}& z_{32}&z_{52}\cr
z_{14}& z_{34}&z_{54}\cr
z_{16}& z_{36}&z_{56}}\ .
\ee
It is convenient to discuss the excitations of the type IIB 
background. 
Begin with the background consisting of $z_{17}$, $q_{15}$, and 
${\tilde p}_5$. After M-reduction along $R_1$ and T-dualization to 
IIB 
along $R_3$, this becomes the canonical F1/NS5/KK-monopole 
background. We 
denote the circle 
dual to the M-theory 13 two-torus by $R_B$, so that the $T^4$ has 
cycles 
$B246$.  Then we can relabel the corresponding charge matrices as:
\be\label{Boddex}
 {\bf Z_o}=\pmatrix{
 D1_{B} & F1_{B} & p_{5} \cr
 D1_{5} & F1_{5} & p_{B} \cr
 N5_{B2465} & D5_{B2465} & D3_{246}} \ ,
\ee
\be\label{Bevenex}
 {\bf Z_e}=\pmatrix{
 {p}_2 & {p}_4 & {p}_6 \cr
 D3_{B46} & D3_{B26} & D3_{B24} \cr
 D3_{546} & D3_{526} & D3_{524}} \ ,
\ee
as well as:
\be\label{Btwoex}
{\bf Z_2}=\pmatrix{F1_{2}& D1_{2}&D3_{5B2}\cr
F1_{4}& D1_{4}&D3_{5B4}\cr
F1_{6}& D1_{6}&D3_{5B6}}\ .
\ee
It is straightforward to find the map of the moduli as well.

As we have seen, the structure of the spectrum of excitations is that 
of
winding/momentum charges on a triplet of independent $T^{4}$'s; 
in addition, there are three special charges. In the matrices above 
the 
${\bf Z_2^t}$ is, entry by entry, the winding dual to the momentum 
excitation ${\bf Z_e}$, for three of the four cycles on the triplet 
of 
$T^4$'s; and: 
\be
 {\bf Z_o}\Gamma=\pmatrix{
 p_{5} & F1_{B} & D1_{B} \cr
 p_{B} & F1_{5} & D1_{5} \cr
 D3_{246} & D5_{B2465} &  N5_{B2465}}~,
\ee
has the remaining three momentum-winding pairs (the pairing is under 
reflection across the diagonal), as well as the three special charges 
on 
the diagonal. The triality of the three $T^4$'s acts by: 
\be
 {\bf Z_e}\rightarrow\Omega {\bf Z_e}\quad,\qquad
  {\bf Z_2}\rightarrow {\bf Z_2}\Omega^{-1}\quad,\qquad
  {\bf Z_o}\Gamma\rightarrow\Omega {\bf Z_o}\Gamma\Omega^{-1}\ ,
\ee
where $\Omega$ is a permutation matrix.  The fact that this operation 
permutes the special charges as well suggests that each of the three 
special 
charges should be associated with a particular $T^4$.

\paragraph{The Spectrum:}
The half-BPS contribution to the masses of the excitations is:
\be
  R_6 E_0=\hf\left(\tr\left[{\bf Z_o} G_o^{-1} {\bf Z_o^t} 
{G}_5\right]
 +\tr\left[{\bf Z_e} G_e^{-1} {\bf Z_e^t} \Gamma G_5^{-1}\Gamma\right]
 +\tr\left[{\bf Z_2} G_o {\bf Z_2^t} G_e\right]\right)\ .
\ee
Specializing this expression to rectangular tori and exploiting the 
fixed 
scalar conditions, we recover appropriate denominators in the mass 
formula.
For generic moduli there is not a meaningful concept
of ``levels'', {\it i.e.} integer denominaters in the mass formula. 
However, it follows from the discussion in section~\ref{glob} that 
other
specializations of the moduli, corresponding to $SL(3,\IZ)$ transforms
of the rectangular tori, likewise give simple denominators. That this
also works out in the present formulae is a consequence of
covariance under duality transformations.

The cross-terms in the mass formula come from 1/4-BPS and higher 
contributions and are independent of the moduli. They are determined 
as 
the expressions involving charges that transform in compatible ways 
and 
reduce appropriately in the special case of rectangular tori. The 
result:
\bea
  R_6 E_\times &=&
	\tr\left[{\bf Z_2}{\bf Q_o^{-1}}{\bf Z_{e}}\right]+
	{\bf Z_o}^{ia}{\bf Z_o}^{jb}\Bigl(\Gamma({\bf 
Q_o^{t}})^{-1}\Bigr)^{ck}
	\epsilon_{ijk}\epsilon_{abc}+\nonumber \\
 &~&\qquad 	+{\rm triality~permutations~of~(o,e,2/5)}~,
\eea
is invariant, independent of moduli, and has the appropriate
denominators in the expected places. Finally, the quantity:
\be
  R_6 E_{\rm spec}=
   \frac{\left(\tr\left[{\bf Q_o^t} \Gamma {\bf Z_o}
   +{\bf Q_e^t}\Gamma {\bf Z_e}
   +{\bf Q_5^t} {\bf Z_2}\right]\right)^2}
	{6\,\left(\det{\bf Q_o}+\det{\bf Q_e}+\det{\bf Q_5}\right)}~,
\ee
is the square of the sum of special charges, {\it i.e.} 
$({\tilde p}_5p_5+n_{B2467}f_5+f_7n_{B2465})$ .
Thus we can write, for example:
\bea
 E_L&=&E_0-2E_\times+E_{\rm spec}~,	\\
 E_R&=&E_0+2E_\times-2E_{\rm spec}\ .	
\eea


\section{Is there an exactly solvable CFT in the moduli space?}
\label{cft}
In the AdS/CFT correspondence, it is extremely useful to find a point 
in 
the moduli space where the CFT is amenable to perturbative treatment, 
or
even better, exactly solvable.  For instance, invariants on the 
moduli 
space, such as the BPS spectrum and its degeneracy, quantum 
corrections 
to current algebra anomalies, etc., can be worked out explicitly. 
Thus 
one may ask whether the three-charge brane background we have been 
discussing admits an exactly solvable CFT in the moduli space.

Let us recall the structure of the symmetric product orbifold CFT
that appears in the moduli space of the two-charge
(\eg\ D1-D5) brane background,
given its remarkable similarity to the present problem.
For $T^4$, the CFT in question is a sigma model on 
$\symn(T^4)\times (\ttil^4\times {\tilde \IR}\times {\tilde S}^3)$,
with $N=q_1q_5$; it describes a region in the moduli
space that is naturally associated to background brane
charges $(q_1,q_5)=(N,1)$.
The other factors in the target space represent the zero modes 
of the diagonal U(1) in the $U(q_1)\times U(q_5)$ gauge theory
on the D1- and D5-branes; the ${\tilde\IR}\times{\tilde S}^3$ 
part, representing motion transverse to the $T^4$, decouples and
may be ignored in this instance.
The CFT moduli space is $\HH_N\backslash SO(5,4)/SO(5)\times SO(4)$,
with $\HH_N$ the discrete subgroup of $SO(5,4;\IZ)$
that preserves the background charge 
vector~\cite{Seiberg:1999xz,lm99}.
BPS considerations~\cite{lm99} and an analysis of the
gravitational effective field theory~\cite{Kutasov:1999xu}
show the existence of a $U(1)^8_L\times U(1)^8_R$
current algebra; half of these come from the diagonal 
of the symmetric orbifold, half from the extra $\ttil^4$.
These currents are a part of the singleton sector of the CFT.
Moduli deformations affect the CFT in two ways: First, they act
as deformations of the $T^4$ components of the symmetric
product orbifold (sixteen of the moduli are the metric and B-field
of the individual $T^4$'s, and four are $\IZ_2$ twist operators);
roughly speaking, the various twisted sectors realize
supergravity in the bulk of $AdS_3\times S^3$.
Secondly, the moduli act in the global, singleton sector 
via current-current interactions which deform the metric 
on the $U(1)^{16}$ charge lattice.  The spectrum of U(1) charged
excitations in the symmetric orbifold, as well as other 
considerations,
is sufficient to identify the symmetric orbifold locus
in a corner of the moduli space that corresponds to canonical 
background
charges $(N,1)$~\cite{lm99}.  Deformations away from
this locus are a linear combination of the moduli of
the symmetric product, and the current-current interactions.

The addition of KK monopoles to the background leads to
a remarkably parallel structure to the moduli space, as we
have seen.  It is therefore tempting to speculate that there
is again an exactly solvable point in the moduli space
in the region corresponding to the canonical background
charges $(N,1,1)$.  The near-horizon geometry of one KK monopole
is the same as flat space; in particular, it
does not deform the angular $S^3$ of the space transverse
to the onebrane-fivebrane system. 
This leads one to suppose that the symmetric
orbifold structure survives more or less intact.

The moduli space is now 
$\HH_N\backslash F_{4(4)}/SU(2)\times USp(6)$.
In addition to the twenty moduli of the D1-D5 system,
there are eight additional moduli; four of these come
from the enlargement of the T-duality group from 
$SO(4,4)$ to $SO(5,4)$ (mixing the fibered circle of
the KK monopole with the four-torus), and four more are RR moduli%
\footnote{When the background charges are entirely NS.}.
The spacetime CFT has twelve right-moving currents in the 
({\bf 2},{\bf 6})
of $SU(2)\times USp(6)$, and fourteen left-moving currents
in the $({\bf 1},{\bf 14})$.  In the IIB description, these currents 
naturally
fall into three sets of $U(1)^4_L\times U(1)^4_R$,
together with two additional `special currents' on the left.
One of these sets of $U(1)^4_L\times U(1)^4_R$
corresponds to perturbative string momentum and winding
(\ref{eq:spec1}), and is naturally associated
with the one-brane background; another set (\ref{eq:spec2}) consists
of D1- and D3-branes, and is naturally associated
with the fivebrane background~\cite{Kutasov:1999xu,lm99}
and the third set (\ref{eq:spec3}) is associated 
to the KK monopole background.
We propose to again identify the first set
of currents with the translation currents
on the diagonal $T^4$ of the symmetric product
$\symn(T^4)$, while the
other two sets are tentatively identified with two
extra four-tori, denoted $\ttil^4$ and $\that^4$
(two (4,0) hypermultiplets are the minimal additional
field content needed to realize these currents).

Fivebrane anomalies~\cite{Maldacena:1997de,Harvey:1998bx}
have terms linear and cubic
in the brane charges.  Since there is no constant term,
the contribution to the anomaly in the SU(2) R-current
coming from the extra $\ttil^4 \times \that^4$
hypermultiplets must be cancelled; the simplest possibility is to
add vectormultiplets, which might be thought of again
as representing zero-modes of the bound state in
the $\IR\times S^3$ throat transverse to the four-torus.

One important problem is to realize the structure of
the moduli space in this framework.  
The 28 moduli of $F_{4(4)}/SU(2)\times USp(6)$
transform as a $({\bf 2},{\bf 14^{\prime}})$ 
under $SU(2)\times USp(6)$.
They again appear in two ways: 
As moduli of the symmetric product orbifold, and as
current-current interactions.
The T-duality group $SO(5,4)$ is manifestly realized in
the symmetric orbifold, when we supplement the $U(1)^4_L\times 
U(1)^4_R$
translation currents on each individual four-torus with the $J_3$
component of the left-moving $SU(2)$ R-symmetry
\footnote{Indeed, it is precisely this quantum number that  
is affected by the asymmetric orbifold which introduces
KK monopole charge in pertubative string 
descriptions~\cite{Giddings:1993wn,Cvetic:1996yq,Kutasov:1998zh}.}.
The four $\IZ_2$ twist moduli that preserve the $\NN=(4,4)$
supersymmetry of the symmetric product may be supplemented
with four more such moduli that only preserve $\NN=(4,0)$
supersymmetry.  These moduli were identified in~\cite{lm99}
by matching quantum numbers in the symmetric
orbifold to those of the supergravity spectrum.
One also needs to represent the effect of the moduli
on the charged spectrum of the singleton sector.
The marginal deformations of a set of U(1) currents
are always of the form $SO(p,q)/SO(p)\times SO(q)$; therefore, the 
$F_{4(4)}/SU(2)\times USp(6)$ local geometry of the moduli space
must, as far as the singleton sector is concerned,
embed in such a structure.
Indeed, $F_{4(4)}$ embeds in $SO(12,14)$ 
via the ${\bf 26}$,
while $SU(2)\times USp(6)$ embeds in $SO(12)$ via the 
$({\bf 2},{\bf 6})$
and in $SO(14)$ via the $({\bf 1},{\bf 14})$.
Note that 
$({\bf 2},{\bf 6})\otimes({\bf 1},{\bf 14})\supset({\bf 
2},{\bf 14^{\prime}})$; there
is a unique projection of the current bilinears
onto the appropriate subspace of $SO(12,14)$.

Clearly $T^4_{\rm diag}\times\ttil^4\times \that^4$ 
realizes an $SO(12,12)$ structure; but one must identify
two more left-moving currents to fill out the ${\bf 14}$ 
of $USp(6)$. From (\ref{eq:eltemp}), one current is (momentum/winding)
of perturbative strings on the KK monopole circle.
The $SO(5,4)$ T-duality group mixes this current with
the analogous ones on $T^4$. 
Under T-duality, one has the decomposition:
\[\begin{array}{ccccc}
  F_{4(4)} & \rightarrow & SO(5,4) & \rightarrow & 
        SO(5)\times SO(4) \\
  {\bf 26} &  & {\bf 9}+{\bf 16}+{\bf 1}& & 
   [({\bf 5};{\bf 1},{\bf 1})+({\bf 1};{\bf 2},{\bf 2})]\,+ \\
	& & & &
   [({\bf 4};{\bf 1},{\bf 2})+({\bf 4};{\bf 2},{\bf 1})]+
   ({\bf 1};{\bf 1},{\bf 1})\ .  \\
\end{array}\]
The natural realization of $SO(5,4)$ on the global
modes of the candidate CFT mixes 
the diagonal $T^4$ translation currents with 
the overall $J_L^3$ of the symmetric product; 
this accounts for the first line in the last column.
The second pair of representations in brackets 
consists of the translation
currents of $\ttil^4\times \that^4$; the remaining singlet 
must be made from some left-moving current, since it is
the last element of the ${\bf 14}$.
Note that the T-duality $SO(5,4)$ naturally embeds in
the $SO(8,8)$ T-duality group of $\ttil^4\times \that^4$
via the spinor representation
(\ie\ the sixteen translation currents transform as the vector
of $SO(8,8)$ and the spinor of $SO(5,4)$).  
This is precisely the way in which D-brane charges transform
under T-duality.
The $SO(5,4)$ structure is thus manifestly realized in the candidate 
CFT.

Modular invariance suggests completion of the signature
(12,14) lattice transforming under $F_{4(4)}$, to a (14,14) lattice
as explained in section~\ref{currents};
the extra charges are the special BPS charge (\ref{eq:spec3}),
and the $SU(2)_R$ R-charge. The latter is natural, since it
contains the right-moving partner of $J_L^3$,
the current of translations on the KK monopole circle.
A puzzling feature of the spectrum derived in section~\ref{currents}
is the difference in left- and right-moving levels,
which is an invariant on the moduli space. The
contribution of the charges to this difference is:
\bea
  Q_R^2-Q_L^2 &=&
	\frac1{q_1}(p^i w_i) +\frac1{q_5}(w^{D1}_i w_{D3}^i)
	+\frac1{q_k}(w_i^{D}w^i_{\tilde D})\nonumber\\
	& &+\frac1{4q_1q_5q_k}\Bigl[J^2-
		(q_kp_5)^2-(q_5w^5)^2-(q_1n_{12345})^2\nonumber\\
	& &\qquad+2(q_5q_k)p_5w^5+2(q_kq_1)p_5n_{12345}
		+2(q_1q_5)w^5n^{12345}\Bigr]~.
\eea
Matching this structure places strong constraints on
the spacetime CFT, and in particular the currents
that remain to be identified in the proposed candidate -- 
the current that couples to the special BPS charge 
on the right, and the last member of the ${\bf 14}$ on the left.

In the $(4,4)$ theory -- the D1/D5-system -- 
the BPS states of the symmetric product orbifold match those
of the KK towers of the effective 6d supergravity
on $AdS_3\times S^3$ 
\cite{Maldacena:1998bw,Larsen:1998xm,deBoer:1998ip}.
The $SU(2)$ R-symmetry current
algebra quantum numbers label the spherical harmonics
on $S^3$.  The 5d supergravity on $AdS_3\times S^2$
that results when KK monopoles are added to the background
is simply the truncation of this spectrum to 
$J_L^3$=0, in the sector with vanishing momentum along the
KK fiber circle.  Deformation in the moduli space
away from the $(N,1,1)$ corner will `squash' the $S^3$,
so that by the time one is in a region with a valid 
low-energy supergravity interpretation, the states carrying
momentum $J_L^3$ on the monopole circle will be much heavier
that those with $J_L^3=0$.  Note that the squashing
of the three-sphere will act as well on the two vectormultiplets
describing the $\IR\times S^3$ throat transverse to the four-torus
parametrized by the hypermultiplets, suggesting that the last element
of the left-moving ${\bf 14}$ should also involve the
bosonic $SU(2)$ WZW currents.

There are other requirements on the spacetime CFT.
One can analyze the F1-NS5-KK monopole background
in global coordinates on $AdS_3$~\cite{Kutasov:1998zh}, using the 
perturbative string techniques of~\cite{Giveon:1998ns}
\footnote{Note that this approach describes the background
only on singular loci, and for $q_k>1$, a rather different
subspace of the moduli space; hence one can only
safely compare invariant quantities such as BPS spectra.}.
At least a subset of the BPS spectrum should
be realized in this framework~\cite{Seiberg:1999xz}.
In~\cite{Kutasov:1998zh}, BPS states carrying the quantum numbers
of perturbative strings with winding and momentum on
the KK monopole circle (parametrized by $x^5$)
were identified; these have dimension $h_R=j_R=j_R^3$ and 
carry oscillator excitation level
$N_{\rm osc}=p_5 w^5$, and thus their degeneracy is 
exponential in $\sqrt{p_5w^5}$.

Thus far, we have not been able to find a suitable
modification of the candidate
$\symn(T^4)\times(\ttil^4\times{\tilde\IR}\times{\tilde S}^3)
\times(\that^4\times{\hat\IR}\times{\hat S}^3)$
CFT meeting all the above requirements.  The crucial
missing ingredient is an identification of the proper
linear combination of the many available currents
which realizes the structure of the `special' charges
$p_5$, $w^5$ and $n_{12345}$.


\paragraph{Comments on K3:}
Finally, we should remark on the differences between 
compactification on $T^4$ and K3.  For D1-D5 bound states on K3, 
there was rather little difference; the extra $\ttil^4$ 
was simply replaced by an extra factor in the symmetric
product, and the duality group was still sufficient to
place all brane charges $(q_1,q_5)$ with the same product
$N=q_1q_5$ within the same moduli space.  The addition
of KK monopoles to the mix yields a difference;
the duality group is no longer sufficient
to map all brane charges to $(N,1,1)$ -- there is no
duality that mixes the KK monopole charge with the other
two, as there is in a $T^4$ compactification.  This is reflected
in the presence of a linear term $q_k(q_1q_5+2)$
in the anomaly of the R-current.  A consequence is that
one expects to have the possibility of an exactly solvable
point in the moduli space only for $q_k=1$.

The 5d supergravity on $AdS_3\times S^2 \times(S^1\times K3)$ 
inherits 22 vectors from tensor multiplets of IIB on K3, two from 
each of the two gravitino multiplets, and one additional vector 
(the graviphoton) from the supergravity 
multiplet~\cite{Cadavid:1995bk}. 
The moduli space is locally $SO(21,4)/SO(21)\times SO(4)$, with the 
charges transforming as 
$({\bf 1},{\bf 21})\oplus ({\bf 2},{\bf 2})\oplus 2({\bf 1},{\bf 1})$ 
under $SO(4)\times SO(21)$.  The first two consist of 
the (4,20) lattice of D1/D3/D5-branes wrapping the monopole
circle and cycles of K3, together with one linear combination
of the special charges; one singlet is the other combination
of special charges coupling to a left-moving current;
and the remaining singlet
is the special BPS charge dual to the brane background,
which couples to a right-moving current.  Adding the 
R-current, the lattice has signature $(6,22)$. The
four multiplets of charges each have components in common with
the toroidal case; so global symmetries, and the discussion of the 
latter
case in this paper, combine to ensure that all the charges 
are associated to currents.



\vskip 1cm

{\bf Acknowledgments:}
We thank 
D. Kutasov 
for helpful discussions.
This work was supported by DOE grant DE-FG02-90ER-40560 and
FL was further supported by a Robert R. McCormick Fellowship. 
FL thanks the theory group at the University of Michigan in Ann Arbor 
and the ITP at the University of California in Santa Barbara for 
hospitality.

\appendix
\section{Derivation of a BPS mass formula}
\label{app:BPS}
The computation in this Appendix follows~\cite{Obers:1998fb}; see
also the Appendix of~\cite{lm99}.

The supersymmetry algebra in M-theory leads to the eigenvalue 
equation for the central charges:
\be
\left[ {\cal C}\Gamma^M P_M + {1\over 2}{\cal C}\Gamma_{MN} Z^{MN}
+{1\over 5!}{\cal C}\Gamma_{MNPQR} Q^{MNPQR}\right]\epsilon = 0~.
\label{eq:susyeq}
\ee
The central charges $Z_{MN}$ and 
$Q^{MNPQR}\equiv {1\over 2}\epsilon^{MNPQRST}Q_{ST}$ are the M2- and 
M5-brane charges; the $P_M$ are the momenta, in particular $P_0$ is 
the 
mass $M$ that we want to compute. The spinorial eigenvector of the 
preserved supersymmetry is denoted $\epsilon$, and the metric is 
mostly plus. 

We choose the background as the three $M5$-branes intersecting over a 
line, and want to consider any configuration of $U(1)$ charges.
Writing out the eigenvalue equation in terms of individual charges
yields an expression that is lengthy and not illuminating. Motivated
by the qualitative considerations in the main text, the charges
can be divided into various groups:
\begin{center}
\begin{tabular}{c|cccccc} 
        level &   $M2$  &  $KK$   &  $M5$  \\ \hline
        $Q_{12}$ & $Z_{35}, Z_{46}, Z_{36}, Z_{45}$ & $P_1, P_2$ & 
                    $Q_{17}, Q_{27}$    \\ 
        $Q_{34}$ & $Z_{15}, Z_{26}, Z_{16}, Z_{25}$ & $P_3, P_4$ & 
                    $Q_{37}, Q_{47}$ \\  
        $Q_{56}$ & $Z_{13}, Z_{24}, Z_{14}, Z_{23}$ & $P_5, P_6$ & 
                    $Q_{57}, Q_{67}$    \\ 
        $Q_{12}+Q_{34}+Q_{56}$ & $Z_{12}, Z_{34}, Z_{56}$ 
&  &       
\end{tabular}
\end{center}
We consider first the terms that correspond to the third line in the 
table, that is a single background $M5$ and $8$ specific $U(1)$ 
charges. 
The eigenvalue equation becomes:
\bea
\lambda_3 &=& Q_{56}\Gamma^{012347} + 
\left[\Gamma^{05}P_5 + \Gamma^{06}P_6
+ \Gamma^{012346}Q_{57} + \Gamma^{012345}Q_{67}\right]  
\nonumber \\
&~&\qquad + \left[\Gamma^{013}Z_{13} +\Gamma^{024}Z_{24}
+ \Gamma^{013}Z_{14} +\Gamma^{013}Z_{23}\right]~,
\eea
which is understood as an operator equation acting on the spinor 
$\epsilon$. 
The three terms on the right hand side are mutually anti-commuting, 
so the 
square of the equation becomes:
\bea
\lambda^2_3 &=& Q_{56}^2 + 
\left[\Gamma^{05}P_5 + \Gamma^{06}P_6
+ \Gamma^{012346}Q_{57} + \Gamma^{012345}Q_{67}\right]^2 
\nonumber \\
&~& \qquad +\left[\Gamma^{013}Z_{13} +\Gamma^{024}Z_{24}
+ \Gamma^{013}Z_{14} +\Gamma^{013}Z_{23}\right]^2 \\
&=&  Q_{56}^2 + (P_5 + Q_{67} )^2 + (P_6 + Q_{57})^2 
\nonumber \\
&~& \qquad +(Z_{13} + Z_{24})^2 + (Z_{14} + Z_{32})^2~.
\eea
In the computation leading to the second expression we took 
$\Gamma^{1234}=1$, thus removing all operators. We can therefore
take an ordinary square root, and then expand the result according
to the hierarchy between the charges. This gives:
\bea
\lambda_3 \simeq Q_{56} &+& {1\over 2Q_{56}}
\left[(P_5 + Z_{67} )^2 + (P_6 + Z_{57})^2 \right. \nonumber \\
&+& \left. (Z_{13} + Z_{24})^2 + (Z_{14} + Z_{32})^2
\right]~.
\label{eq:lam1}
\eea
An analogous computation can be carried out for the charges in the
first and second line of the table, yielding eigenvalues $\lambda_1$ 
and 
$\lambda_2$, respectively, after imposing the conditions 
$\Gamma^{3456}= 1$ and 
$-\Gamma^{1256}=\Gamma^{1234}\Gamma^{3456}=1$.  
Now note that the terms 
arising from any of the first three lines of the table commute with 
those coming from the others, in particular $\Gamma^{1234}$, 
$\Gamma^{3456}$, and $\Gamma^{1256}$ mutually commute.
Therefore the involved matrices are simultaneously diagonalisable or, 
in other words, the full result is simply the sum of three terms of 
the form (\ref{eq:lam1}). 

At this point we need to include also the three ``special'' $U(1)$ 
charges that appear in the fourth line of the table. Let us consider 
these 
together
with the background, without other $U(1)$ charges present. The 
eigenvalue
problem becomes\footnote{The signs of $Q_{34}$ and $Z_{34}$
have been flipped relative to (\ref{eq:susyeq}) so
that the background charges can be taken positive. Below $Q_{37}$ and
$Q_{47}$ will be similarly flipped.}:
\be
\mu = 
Q_{12}\Gamma^{034567}- Q_{34}\Gamma^{012567} +Q_{56}\Gamma^{012347}
+ Z_{12}\Gamma^{012} - Z_{34}\Gamma^{034}+ Z_{56}\Gamma^{056}~.
\ee
The first three and the last three terms commute between themselves,
but these two groups anticommute with each other. The square therefore
gives:
\bea
\mu^2 =  (Q_{12}+ Q_{34} + Q_{56})^2
+ (Z_{12} +  Z_{34} + Z_{56})^2~,
\eea
for $\Gamma^{1234}=\Gamma^{3456}= - \Gamma^{1256}=1$. 
This is an exact expression for the mass, when only the background and
the three special $U(1)$ charges are turned on. In the scaling limit
it becomes:
\bea
\mu \simeq Q_{12}+ Q_{34} + Q_{56}
+ {(Z_{12} +  Z_{34} + Z_{56})^2 \over 
2(Q_{56}+ Q_{12} + Q_{34})}~.
\eea

The transformations that diagonalize the eigenvalue problems for
$\lambda_{1,2,3}$ commute with each other, as remarked above, 
but they do not in general have a simple relation to the 
eigenvalue problem for $\mu$. However, to the order we compute
the eigenvectors of the undisturbed background solve all the 
eigenvalue problems considered; and this is sufficient to 
guarantee that the various partial results can be added without
inducing further crossterms. The final result for the mass is 
therefore:
\bea
M &\simeq& Q_{12}+ Q_{34} + Q_{56} \nonumber \\
&+& {1\over 2Q_{12}}\left[(P_1 + Q_{27} )^2 + (P_2 + Q_{17})^2 + 
(Z_{35} + Z_{46})^2 + (Z_{36} + Z_{54})^2 \right] \nonumber \\
&+& {1\over 2Q_{34}}\left[(P_3 + Z_{47} )^2 + (P_4 + Q_{37})^2 + 
(Z_{15} + Z_{62})^2 + (Z_{16} + Z_{25})^2 \right] \nonumber \\
&+& {1\over 2Q_{56}}\left[(P_5 + Q_{67} )^2 + (P_6 + Q_{57})^2 + 
(Z_{13} + Z_{24})^2 + (Z_{14} + Z_{32})^2 \right] \nonumber \\
&+& {(Z_{12} +  Z_{34} + Z_{56})^2 \over 
2(Q_{12}+ Q_{34} + Q_{56})}~.
\label{eq:mas}
\eea
This is the formula needed in the main text.

In the considerations relating to black hole entropy we need a couple 
of 
refinements. First, it is customary to include the scalar charge,
{\it i.e.} the momentum $P_7$ 
along 
the $R_7$ direction. The resulting  term in the eigenvalue equation 
commutes with all the background terms and anticommuetes with all the
terms from the $U(1)$ charges. Additionally, the $P_7$ is 
small in the scaling limit, of order $l_p^0$, so cross-terms between 
the
scalar charge and the $U(1)$ charges are negligible.  These facts are 
sufficient to ensure that $P_7$ can be included in (\ref{eq:mas}) by 
simple addition.  

Next, we want to derive the conformal weight of the left-movers.
Roughly, this amounts to the substitution 
$\epsilon\to\Gamma^{7}\epsilon$,
flipping the eigenvalue of the momentum; alternatively one might flip 
the five-brane numbers by taking the opposite quantum numbers for
$\Gamma^{1234}$, $\Gamma^{3456}$, and $\Gamma^{1256}$. Either way 
there is a problem, because the eigenvalues of the $\Gamma$-matrices 
are not independent. The physical origin of the problem is that the 
present
theory is chiral, with $(4,0)$ supersymmetry. The conformal weight of 
the 
left-movers is therefore {\it not} given by supersymmetry alone. This
contrasts with the treatment of the $D1-D5$ system in~\cite{lm99},
where the $(4,4)$ supersymmetry was exploited to deduce both the 
right-moving and the left-moving weights. In the main text we find 
the left-moving weights using global symmetries.


\end{document}